\documentstyle[aps,prd,preprint,floats,epsf]{revtex}
\begin{document}
\draft

\tighten 
\preprint{\tighten\vbox{
              \hbox{\hfil CLNS 99/1635}
              \hbox{\hfil CLEO 99--13}
              \hbox{\hfil 20 October 1999 \vspace{6pt}}
}}


\title{Hadronic Structure 
       in the Decay \boldmath $\tau^-\rightarrow\pi^-\pi^0\nu_\tau$}

 
\author{The CLEO Collaboration\vspace{12pt}}
\date{Submitted to Phys.~Rev.~D}

\maketitle


\begin{abstract} 
We report on a study of the invariant mass spectrum of the hadronic 
system in the decay $\tau^-\to \pi^-\pi^0\nu_\tau$.  
This study was performed with data obtained with the CLEO~II 
detector operating at the CESR $e^+e^-$ collider.  
We present fits to phenomenological models in which 
resonance parameters associated with the $\rho(770)$ and 
$\rho(1450)$ mesons are determined.  
The $\pi^-\pi^0$ spectral function inferred from the 
invariant mass spectrum is compared 
with data on $e^+e^-\to\pi^+\pi^-$ as a 
test of the Conserved Vector Current theorem.  
We also discuss the implications of our data with regard to 
estimates of the hadronic contribution to the muon anomalous 
magnetic moment.  
\end{abstract}





\clearpage

\begin{center}
S.~Anderson,$^{1}$ V.~V.~Frolov,$^{1}$ Y.~Kubota,$^{1}$
S.~J.~Lee,$^{1}$ R.~Mahapatra,$^{1}$ J.~J.~O'Neill,$^{1}$
R.~Poling,$^{1}$ T.~Riehle,$^{1}$ A.~Smith,$^{1}$
S.~Ahmed,$^{2}$ M.~S.~Alam,$^{2}$
% ...
\end{center}
 
\small
\begin{center}
$^{1}${University of Minnesota, Minneapolis, Minnesota 55455}\\
$^{2}${State University of New York at Albany, Albany, New York 12222}\\
% ...
\end{center}

\clearpage

\section{INTRODUCTION}
\label{s-intro}

The $\tau$ is the only lepton heavy enough to decay to final states 
containing hadrons.  Since leptons do not participate in the strong 
interaction, $\tau$ lepton decay is well-suited for isolating 
the properties of hadronic systems produced via the hadronic weak 
current~\cite{Tsai,ThackSak}. Furthermore, angular momentum 
conservation plus the transformation properties under parity and 
G-parity of the vector and axial vector parts of the weak current 
give rise to selection rules that constrain the types of hadronic 
states that may form.  
Thus, $\tau$ lepton decay provides an especially clean environment 
for studying these states.  
In this article, we present a 
study of the $\pi^-\pi^0$ system~\cite{chrg} produced in the decay 
$\tau^-\to\pi^-\pi^0\nu_\tau$ based on data collected with the 
CLEO~II detector.  

In semi-hadronic $\tau$ decay, hadronic states 
consisting of two pseudoscalar mesons may only 
have spin-parity quantum numbers $J^P = 0^+$ or $1^-$.  
In addition, the Conserved Vector Current theorem (CVC)
forbids production of $0^+$ non-strange states in $\tau$ decay.  
Thus, within the picture of resonance dominance in the 
accessible range of squared momentum transfer $q^2$, the decay 
$\tau^-\to\pi^-\pi^0\nu_\tau$ is expected to be dominated 
by production of the lowest lying vector meson, the $\rho(770)$.  
Radial excitations, such as the $\rho(1450)$ and the $\rho(1700)$,  
may also contribute.  Although these are well-known mesons, their 
properties have not been measured precisely, and there exists a 
wide variety of models that purport to characterize their line 
shapes.  New data can help improve the understanding of these 
states.  

Finally, CVC relates properties of the $\pi^-\pi^0$ system produced 
in $\tau$ decay to those of the $\pi^+\pi^-$ system produced in the 
reaction $e^+e^-\to\pi^+\pi^-$ in the limit of exact isospin symmetry. 
The degree to which these relations hold has important consequences.  
For example, data on the $e^+e^-$ process is used to determine the 
dominant contribution to 
the large but uncalculable hadronic vacuum-polarization 
radiative corrections to the muon anomalous magnetic moment 
$a_\mu = (g_\mu -2)/2$.  With CVC, $\tau$ data can 
be used to augment the $e^+e^-$ data, leading to a more precise 
Standard Model prediction for the value of $a_\mu$~\cite{adh}.

Here, we attempt to address some of these issues, using 
a high-statistics, high-purity sample of reconstructed 
$\tau^-\to\pi^-\pi^0\nu_\tau$ decays.  The measurements presented 
here supercede earlier preliminary results from CLEO~II on 
this subject~\cite{ju}.  Work in this area has also been 
published by the ALEPH Collaboration~\cite{aleph}.  
In Sec.~\ref{s-models}, we review models of the hadronic 
current in the decays of the $\tau$ to vector mesons,
and specify the models we employ to extract resonance 
parameters.
In Sec.~\ref{s-evtsel} we discuss our data sample
and the event selection criteria.  To mitigate experimental 
biases, we apply several corrections to the data, described in 
Sec.~\ref{s-corr}.  
The results of fits to the corrected $q^2$ spectrum 
are reported in Sec.~\ref{s-results}, and systematic errors 
are discussed in Sec.~\ref{s-syserr}.  We compare our data 
with those obtained by ALEPH and the low-energy $e^+e^-$ 
experiments in Sec.~\ref{s-compexp}.  In Sec.~\ref{s-amu} we 
discuss the applicability of our data for predictions of the 
muon anomalous magnetic moment.  Finally, we summarize 
our results in Sec.~\ref{s-summary}.  

\section{PHENOMENOLOGY AND MODELS}
\label{s-models}

\subsection{Model-Independent Phenomenology}

The decay rate for $\tau^-\to\pi^-\pi^0\nu_\tau$ can be written 
as~\cite{Tsai} 
\begin{equation}
   \frac{d\Gamma(\tau^-\to \pi^-\pi^0\nu_\tau)}{dq^2} = 
   \frac{G_F^2\, |V_{ud}|^2\,S^{\pi\pi}_{EW}}
        {32\pi^2\, M_\tau^3}\,
   (M_\tau^2-q^2)^2 \,
   (M_\tau^2 + 2 q^2 ) \,
    v^{\pi\pi^0}(q^2), 
   \label{eq:dgam} 
\end{equation}
where $q^2$ is the invariant mass squared of the $\pi^-\pi^0$ system,
and $v^{\pi\pi^0}(q^2)$ is the vector spectral function characterizing 
the ({\sl a priori} unknown) 
hadronic physics involved in the formation of the $(J^P = 1^-)$ 
$\pi^-\pi^0$ system.  
$G_F$ is the Fermi constant, $V_{ud}$ the 
Cabibbo-Kobayashi-Maskawa (CKM) matrix element, and $M_\tau$ the $\tau$ 
lepton mass.  
$S^{\pi\pi}_{EW}$ denotes electroweak radiative corrections not 
already absorbed into the definition of $G_F$, some components of 
which have been determined theoretically~\cite{MS,BL,BNP}.

The corresponding $\pi^+\pi^-$ spectral function $v^{\pi\pi}(q^2)$
can be inferred from the cross section for 
$e^+ e^- \to \pi^+\pi^-$ \cite{Tsai,GM}:
\begin{equation}
 \sigma(e^+ e^- \to \pi^+\pi^-)  =  
   \left(\frac{4\pi^2\alpha_{em}^2}{s} \right) v^{\pi\pi}(s) \, , 
   \label{eq:cross}
\end{equation}
where $s = q^2$ is the squared $e^+e^-$ center-of-mass energy.  
Up to isospin-violating effects, 
CVC allows one to relate the spectral function obtained from 
$\tau$ decay to the isovector part of the $e^+e^-$ spectral function:
\begin{equation}
  v^{\pi\pi}_{I=1}(q^2) = v^{\pi\pi^0}(q^2)\, .
\end{equation}
%

The $e^+e^-$ spectral function can also be expressed in terms of 
the pion electromagnetic form factor $F_{\pi}(q^2)$: 
\begin{equation}
   v^{\pi\pi}(q^2) = \frac{1}{12\pi}
                 \left| F_\pi(q^2) \right|^2 
                 \left(\frac{2 p_\pi}{\sqrt{q^2}}\right)^3 , 
   \label{eq:fpi}
\end{equation}
where the last factor represents the $P$-wave phase space factor,
with $p_\pi$ being the momentum of one of the pions in the $\pi\pi$ 
rest frame.  The $\tau$ decay spectral function can be similarly 
expressed in terms of the weak pion form factor.

\subsection{Models of the Hadronic Current}

The hadronic physics is contained within $v^{\pi\pi}(q^2)$, or 
equivalently $F_\pi(q^2)$.  From the 
electric charge of the $\pi^-$, it is known that $F_\pi(0) = 1$.  Beyond 
that, its form at low energies is not presently calculable in QCD, 
and models must be used.  With resonance dominance, it 
is expected that $F_\pi$ is dominated by the line shape of the $\rho(770)$ 
meson, with contributions from its radial excitations, the $\rho(1450)$ 
and $\rho(1700)$ mesons (denoted as $\rho^\prime$ and $\rho^{\prime\prime}$, 
respectively).  

Various Breit-Wigner forms have been 
proposed~\cite{Tsai,GS,pisut,KS,ben} to parameterize $F_\pi$.  
We consider here two models: those of K\"uhn and Santamaria~\cite{KS} and 
Gounaris and Sakurai~\cite{GS}, denoted as the K\&S and G\&S models, 
respectively.  

\subsubsection{The Model of K\"uhn and Santamaria}

In addition to its simplicity, the K\&S model is useful since it is 
implemented in the {\tt TAUOLA} $\tau$ decay package~\cite{korb} used 
in the CLEO~II Monte Carlo simulation.  The form is given by 
\begin{equation}
        F^{(I=1)}_{\pi}(q^2) 
                =   \frac{1}{1+\beta+\gamma+\cdots}\, 
                        \left( BW_\rho + \beta \, BW_{\rho^\prime}
                        + \gamma\, BW_{\rho^{\prime\prime}} +\cdots\right), 
\label{eq:vks}
\end{equation}
where 
\begin{equation} 
   BW_\rho = \frac{M_\rho^2}{(M_\rho^2-q^2)-i\sqrt{q^2}\Gamma_{\!\rho}(q^2)} 
\label{eq:bwks}
\end{equation}
represents the Breit-Wigner function associated with the $\rho(770)$ resonance 
line shape, with $M_\rho$ and $\Gamma_{\!\rho}(q^2)$ denoting the $\rho$ 
meson mass and mass-dependent total decay width.  The assumed 
form for the latter is described below.  
The parameters $\beta$ and $\gamma$ specify the relative 
couplings to $\rho^\prime$ and $\rho^{\prime\prime}$, and the ellipsis 
indicates the possibility of additional contributions.  The Breit-Wigner 
functions are individually normalized 
so that the condition $F_\pi(0) = 1$ is satisfied with 
the inclusion of the $1/(1+\beta+\gamma)$ factor.  
For application to $e^+e^-\to\pi^+\pi^-$ data, 
one must consider isoscalar as well as isovector contributions.  
For this, the form for $F_\pi^{(I=0,1)}$ is obtained by modifying 
$BW_\rho$ so as to characterize $\rho$-$\omega$ interference, 
which is not relevant for $\tau$ decay.  

An alternate form for $v^{\pi\pi}(q^2)$ can be obtained from consideration 
of the amplitudes for weak production and strong decay of $\rho$ mesons.  
For the case where only the $\rho(770)$ contributes, 
the spectral function can be 
expressed (following Tsai~\cite{Tsai}) as:
\begin{equation}
        v^{\pi\pi}(q^2) = 
        \frac{2\pi f_\rho^2}{q^2}
        \left[ \frac{\sqrt{q^2}\,\Gamma_{\!\rho}(q^2)/\pi}
               {(M_\rho^2 - q^2)^2 + q^2\, {\Gamma^2_{\!\rho}(q^2)}}
        \right], 
\label{eq:bw}
\end{equation}
where 
\begin{equation}
\Gamma_{\!\rho}(q^2)= \left( \frac{g_\rho^2}{48\pi}\right)\,
                      \sqrt{q^2}\,
                      \left( \frac{2p_\pi}{\sqrt{q^2}} \right)^3
\label{eq:gamma}
\end{equation} 
gives the energy dependence of the $\rho$ width.  The 
constants $f_\rho$ (with units of mass squared) 
and $g_\rho$ (dimensionless) can be identified as the weak and 
strong $\rho$ meson decay constants, respectively.  
The $F_\pi(0)=1$ condition is satisfied for 
$f_\rho g_\rho = \sqrt{2}M_\rho^2$, in which case the K\&S form is 
recovered.  

The energy dependence of the $\rho$ width may be more complicated than 
the P-wave behavior indicated in Eq.~\ref{eq:gamma}.  
Various authors~\cite{pisut,CD} suggest the need for an additional 
Blatt-Weisskopf centrifugal barrier factor~\cite{BW} which takes the form 
\begin{equation}
 F_R = \frac{1 + R^2 p_0^2}{1 + R^2 p_\pi^2}, 
\label{eq:blatt}
\end{equation}
where $p_0 = p_\pi(q^2 = M_\rho^2)$, and $R$ denotes the range parameter  
with a value assumed to be of ${\cal O}$(1~fermi/$\hbar c$).  This factor 
multiplies the right hand side of Eq.~\ref{eq:gamma}, and thus modifies 
the $\Gamma(q^2)$ factors appearing in both the numerator and denominator 
of the Breit-Wigner form for $v^{\pi\pi}(q^2)$ given by Eq.~\ref{eq:bw}.

\subsubsection{The Model of Gounaris and Sakurai}

The G\&S model~\cite{GS} has been used by a 
number of authors~\cite{aleph,GS,KS,barkov} to parameterize the 
$e^+e^-\to\pi^+\pi^-$ cross section.  In this model, the form for 
$F_\pi$ is derived from an assumed effective range formula for the P-wave 
$\pi$-$\pi$ scattering phase shift, assuming $\rho(770)$ meson dominance.  
This yields 
\begin{equation}
  F_\pi(q^2) = \frac{M_\rho^2 + d\,M_\rho\Gamma_\rho}
                  {(M_\rho^2 - q^2) + f(q^2) -i\sqrt{q^2}\Gamma_\rho(q^2)}, 
\label{eq:gs}
\end{equation}
where $\Gamma_\rho$ denotes $\Gamma_\rho(q^2=M_\rho^2)$, and 
\begin{eqnarray}
  f(q^2) & = & p_\pi^2(q^2)\, \left[ h(q^2) - h(M_\rho^2) \right]
       - p_0^2\, (q^2 - M_\rho^2) \left. 
         \frac{dh}{dq^2}\right|_{\scriptstyle q^2=M_\rho^2} \\
  h(q^2) & = & \frac{\Gamma_\rho M_\rho^2}{p_0^3}\,
                   \frac{2p_\pi(q^2)}{\pi\sqrt{q^2}}\,
                   \ln\frac{\sqrt{q^2} + 2p_\pi(q^2)}{2\,M_\pi}, 
\end{eqnarray}
and $d$ is chosen so as to satisfy the $F_\pi(0)=1$ condition,
\begin{equation}
  d = \frac{3 \,M_\pi^2}{\pi\,p_0^2}\,
      \ln\frac{M_\rho + 2p_0}{2M_\pi}
    + \frac{M_\rho}{2\pi\, p_0}
    - \frac{M_\pi^2M_\rho}{\pi\,p_0^3}.
\end{equation}
Following Refs.~\cite{aleph,KS,barkov}, we employ an extension of this 
model to include possible $\rho^\prime$ and $\rho^{\prime\prime}$ 
contributions, as in Eq.~\ref{eq:vks}. 

The G\&S form for $F_\pi$ is similar to the K\&S form in that 
(1) both are normalized so that $F_\pi(0)=1$, and (2) their shapes 
are similar in the vicinity of the $\rho$ peak, since $f(q^2)$ in 
Eq.~\ref{eq:gs} goes as $M_\rho^2-q^2$ near $q^2=M_\rho^2$~\cite{GS}.  
However, the additional term in the numerator of Eq.~\ref{eq:gs} results in 
a larger value for $F_\pi$ at $q^2=M_\rho^2$ relative to that in the K\&S 
model, given the same values for $M_\rho$ and $\Gamma_\rho$.  
For $M_{\rho}=0.775$~GeV, the value of $d$ is 0.48, 
such that $F_\pi(M_\rho^2)$ is larger by $9\%$ than 
the corresponding value from the K\&S model.

\section{DATA SAMPLE AND EVENT SELECTION}
\label{s-evtsel}

\subsection{Detector and Data Set}

The analysis described here is based on 3.5 fb$^{-1}$ of $e^+e^-$ collision 
data collected at center-of-mass energies  
$2E_{\rm beam}$ of $\sim 10.6$ GeV, 
corresponding to $3.2\times 10^{6}$ interactions of the type 
$e^+e^-\to \tau^+\tau^-(\gamma)$.  These data were recorded at the 
Cornell Electron Storage Ring (CESR) 
with the CLEO~II detector~\cite{cleonim} between 1990 and 1994.  
Charged particle tracking in CLEO~II 
consists of a cylindrical six-layer straw tube array
surrounding a beam pipe of radius 3.2~cm that encloses 
the $e^+e^-$ collision region, 
followed by two co-axial cylindrical drift chambers of 10 and 51 
sense wire layers respectively.  
Scintillation counters 
used for triggering and time-of-flight measurements 
surround the tracking chambers.  For electromagnetic calorimetry, 
7800 CsI(Tl) crystals are arrayed in projective 
and axial geometries in barrel and end cap sections, 
respectively.  The barrel crystals present 16 radiation lengths to 
photons originating from the interaction point.  

Identification of $\tau^-\to \pi^-\pi^0\nu_\tau$ decays relies 
heavily on the segmentation and energy resolution of the calorimeter for 
reconstruction of the $\pi^0$.  The central portion of the 
barrel calorimeter ($|\cos\theta|<0.71$, where $\theta$ is the polar 
angle relative to the beam axis) 
achieves energy and angular resolutions of 
$\sigma_E/E\,(\%) = 0.35/E^{0.75} + 1.9 - 0.1\,E$ and 
$\sigma_\phi\,\mbox{(mrad)} = 2.8/\sqrt{E} + 2.5$, with $E$ in GeV, 
for electromagnetic showers.  The angular resolution ensures that 
the two clusters of energy deposited by the photons from a $\pi^0$ 
decay are resolved over most of the range of $\pi^0$ energies typical 
of the $\tau$ decay mode studied here.

The detector elements described above are immersed in a 1.5 Tesla 
magnetic field provided by a superconducting solenoid surrounding the 
calorimeter.  Muon identification is accomplished with plastic streamer 
tubes, operated in proportional mode, 
embedded in the flux return steel at depths corresponding to 
3, 5 and 7 interaction lengths of total material penetration 
at normal incidence. 

\subsection{Monte Carlo samples}
\label{ss-mc}

We have generated large samples of Monte Carlo (MC) events for use in this 
analysis.  
The physics of the $\tau$-pair production and decay 
is modelled by the {\tt KORALB/TAUOLA} event generator~\cite{korb}, 
while the detector response is handled with a {\tt GEANT}-based~\cite{geant}
simulation of the CLEO~II detector.  The primary MC sample,  
denoted as the generic $\tau$ MC sample, consists of 
11.9 million $\tau$-pair events with all decay modes present.  We  
generated an additional sample enriched in $\tau^-\to\pi^-\pi^0\nu_\tau$ 
decays, bringing the total number of MC signal decays to 10.9 million, 
corresponding to roughly seven times the integrated luminosity of the data.  
The generic $\tau$ Monte Carlo sample is used to estimate backgrounds 
from non-signal $\tau$ decays, as well as for comparisons of kinematic and 
detector-related distributions with those from the data.
We employ the full MC sample for the bin migration 
and acceptance corrections described in Sec.~\ref{s-corr}.

The $\pi^-\pi^0$ spectral function implemented in the Monte Carlo 
is the K\&S model, 
with parameters $(M_\rho,\ \Gamma_{\!\rho},\ \beta,\ M_{\rho^\prime},\ 
\Gamma_{\rho^\prime}) = (0.773,\ 0.145,\ -0.145,\ 1.370,\ 0.510)$   
in GeV, except for $\beta$ which is dimensionless.  
Here $\Gamma_{\!\rho}$ denotes the pole 
$\rho$ meson width, $\Gamma_{\!\rho}(q^2 = M_\rho^2)$.   
These parameters are based on one of the fits by 
K\"uhn and Santamaria~\cite{KS} to the $e^+e^-\to\pi^+\pi^-$ data.  
This fit did not allow for a possible $\rho(1700)$ contribution.  

\subsection{Event selection}

Tau leptons are produced in pairs in $e^+e^-$ collisions.
At CESR beam energies, the decay products of the $\tau^+$ and 
$\tau^-$ are well separated in the CLEO detector.  
The decay of the $\tau^-$ lepton into $\pi^-\pi^0\nu_\tau$ 
is referred to as the signal decay, while that of 
the recoiling $\tau^+$ is referred to as the tag decay, and 
similarly for the charge conjugate case.  
Due to limited charged $\pi/K$ separation capabilities, 
we do not attempt to distinguish $\pi^-$ from $K^-$
in this analysis.  As a result, our selected event sample 
contains background from the Cabibbo-suppressed channel 
$\tau^-\to K^-\pi^0\nu_\tau$.  This and misidentified decays 
from other channels are subtracted statistically using the 
generic $\tau$ Monte Carlo sample described above.

To reject background from non-$\tau\tau$ events, 
we require the tag decay products 
to be identified with one of three 
decay channels: $e^+ \nu_e \overline{\nu}_\tau$ (``$e$ tag''),
$\mu^+ \nu_\mu \overline{\nu}_\tau$ (``$\mu$ tag''),
and $\pi^+\pi^0\overline{\nu}_\tau$ (``$\rho$ tag'').
For the ``$\rho$~vs.~$\rho$'' topology, each event is considered 
twice, corresponding to the two ways of labelling the 
decays as tag and signal decays.  Thus, in such events 
both decays are used in our analysis if the requirements 
given below are met for both combinations of tag and signal labels. 
We have previously used these event topologies to measure the 
branching fraction for the signal decay mode, 
described in Ref.~\cite{cleorho}. 
The event selection used here is similar and is described below.  

We require an event to contain exactly 
two reconstructed charged tracks, separated in angle by 
at least $90^\circ$.  Both tracks must lie in the central 
region of the detector: the tag track must lie within 
$|\cos\theta|<0.8$, while the signal track must have 
$|\cos\theta|<0.71$, so as to avoid excess interactions 
in the main drift chamber end plate.  
Both tracks must be consistent with originating from the 
$e^+e^-$ interaction region, 
and have momentum between $0.08\, E_{\rm beam}$ and 
$0.90\, E_{\rm beam}$.  The momenta of all charged tracks 
are corrected for dE/dx energy loss in the beam pipe and 
tracking system.  

Clusters of energy deposition in the calorimeter are considered 
as candidates for photons from $\pi^0$ decay if they are observed
in the central part of the detector ($|\cos\theta|<0.71$),
are not matched to a charged track, 
and have energy greater than 50 MeV.  
Pairs of photons with invariant mass $M_{\gamma\gamma}$ 
within 7.5 $\sigma_{\gamma\gamma}$ of the $\pi^0$ mass 
are considered as $\pi^0$ candidates.  
The $\gamma\gamma$ invariant mass resolution 
$\sigma_{\gamma\gamma}$ 
varies from 4 to 7 MeV/c$^2$, depending on $\pi^0$ energy 
and decay angle.  
The $\pi^0$ energy is required to be greater than
$0.08\, E_{\rm beam}$.  
Each $\pi^0$ candidate is associated with the charged track
nearest in angle to form a $\pi^-\pi^0$ candidate.
If more than one $\pi^0$ candidate can be assigned 
to a given track, only one combination is chosen,  
namely that for which the largest unused barrel photon-like 
cluster in the $\pi^-\pi^0$ hemisphere has the least energy.  
A cluster is defined to be photon-like if it 
has a transverse energy profile consistent with 
expectations for a photon, 
and if it lies at least 30~cm away from the nearest track projection.  

As mentioned earlier, backgrounds from multihadronic 
($e^+e^- \to q\overline q$) events 
are rejected by identifying the tag system 
as being consistent with $\tau^+$ 
decay to neutrino(s) plus $e^+$, $\mu^+$ or $\pi^+\pi^0$. 
The tag track is identified as an electron if its calorimeter 
energy to track momentum ratio satisfies $0.85<E/p<1.1$ and if its 
specific ionization in the main drift chamber is no more than two 
standard deviations ($\sigma$) 
below the value expected for electrons.  
It is classified as a muon if the track has penetrated at least the 
innermost layer of muon chambers at 3 interaction lengths.  
If the tag track is not identified as an $e$ or a $\mu$, but is 
accompanied by a second $\pi^0$ of energy $\ge 350\,{\mbox MeV}$, 
then the track-$\pi^0$ combination is classified as a $\rho$ tag.  
The invariant mass of this combination must be between 0.55 and 1.20 GeV.

To ensure that these classifications are consistent with 
expectations from $\tau$ decay, events are vetoed if any unused 
photon-like cluster with $|\cos\theta| < 0.95$ 
has energy greater than 100 MeV, or if any 
unmatched non-photon-like cluster has energy above 500 MeV.  Finally, 
the missing momentum as determined using the $\pi^-\pi^0$ and tagging 
systems must point into a high-acceptance region of the detector 
($|\cos\theta_{\rm miss}| < 0.85$), and must have a component 
transverse to the beam of at least $0.08\, E_{\rm beam}$.  These 
requirements also limit the misidentification of $\tau$ decays 
containing multiple $\pi^0$'s as signal decays.  

\subsection{Final event sample}
\label{ss-evtsample}

\begin{figure}
   \leavevmode\centering 
   \epsfxsize=6.5in
   \epsfbox{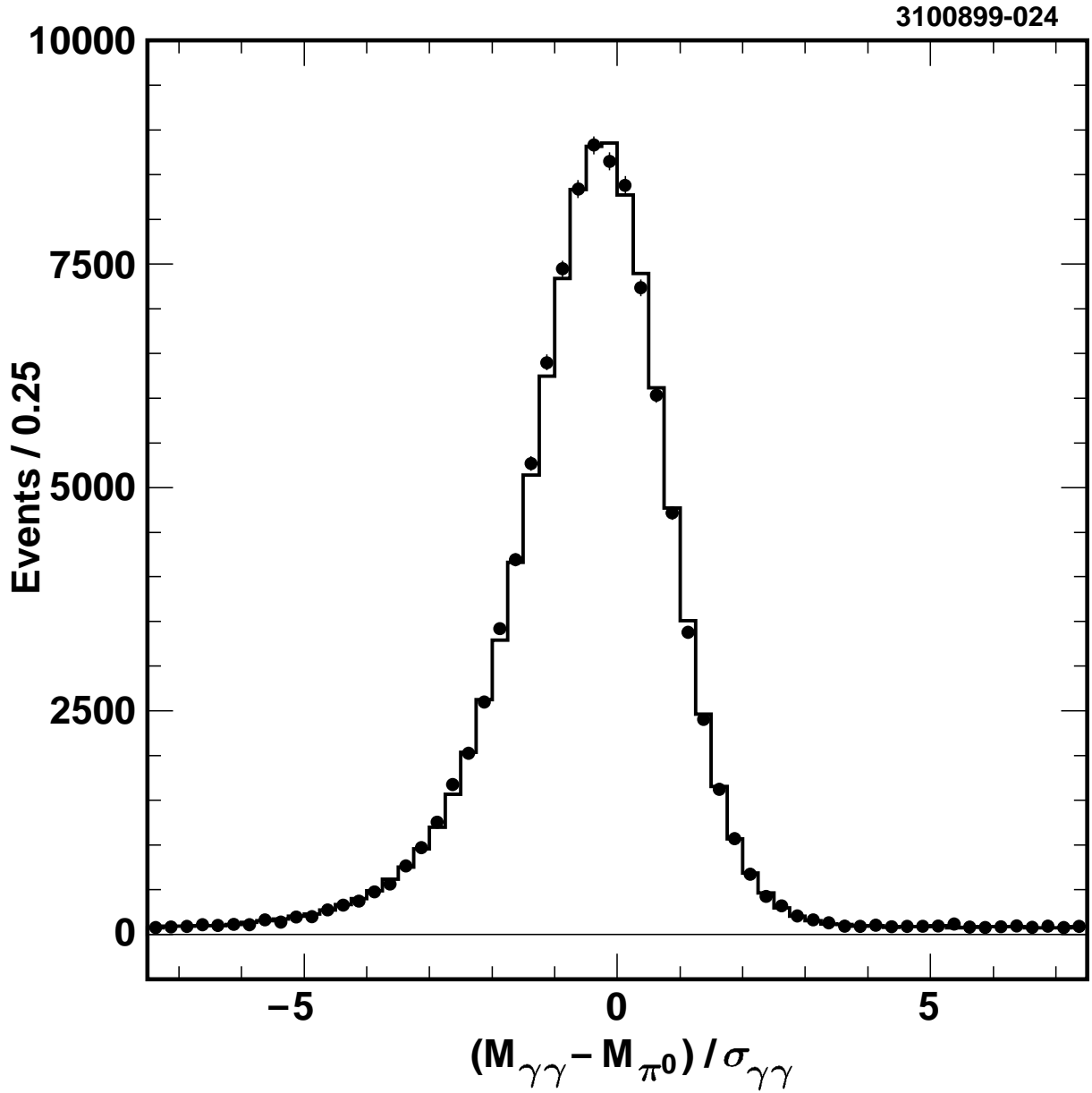}
   \caption{Normalized $M_{\gamma\gamma}$ distribution for $\pi^0$ 
            candidate photon pairs in the data (points) and the 
            generic $\tau$ Monte Carlo sample (line histogram), 
            after all other cuts.  
           }
  \label{fig:spi0}
\end{figure}

With this selection, 103522 events remain.  
The distribution in normalized di-photon invariant mass 
$S_{\gamma\gamma} = (M_{\gamma\gamma}-M_{\pi^0})/\sigma_{\gamma\gamma}$ 
for these events is shown in Fig.~\ref{fig:spi0}, with 
the corresponding Monte Carlo distribution overlaid.  
%
%
Of these, 94948 lie in the $\pi^0$ signal region, defined to be the 
interval $-3.0 < S_{\gamma\gamma} < 2.0$.  The asymmetry of the distribution 
and the signal region definition arises because of the asymmetric 
energy response of the calorimeter.  The low-side tail of the photon energy 
response curve is due primarily to rear and transverse leakage of high energy 
showers out of the CsI crystals whose energy depositions are summed 
in determining the energy of a given photon.  We also make use of 2281 
events lying in the side-band regions $-7.5 < S_{\gamma\gamma} < -5.0$ 
and $3.0 < S_{\gamma\gamma} < 5.5$ to model backgrounds associated with 
spurious $\pi^0$ candidates.  After these selections, we redetermine 
the photon energies 
and angles making use of the $\pi^0$ mass constraint, so 
as to improve the $\pi^-\pi^0$ invariant mass resolution.

\begin{figure}
   \leavevmode\centering 
   \epsfxsize=6.5in
   \epsfbox{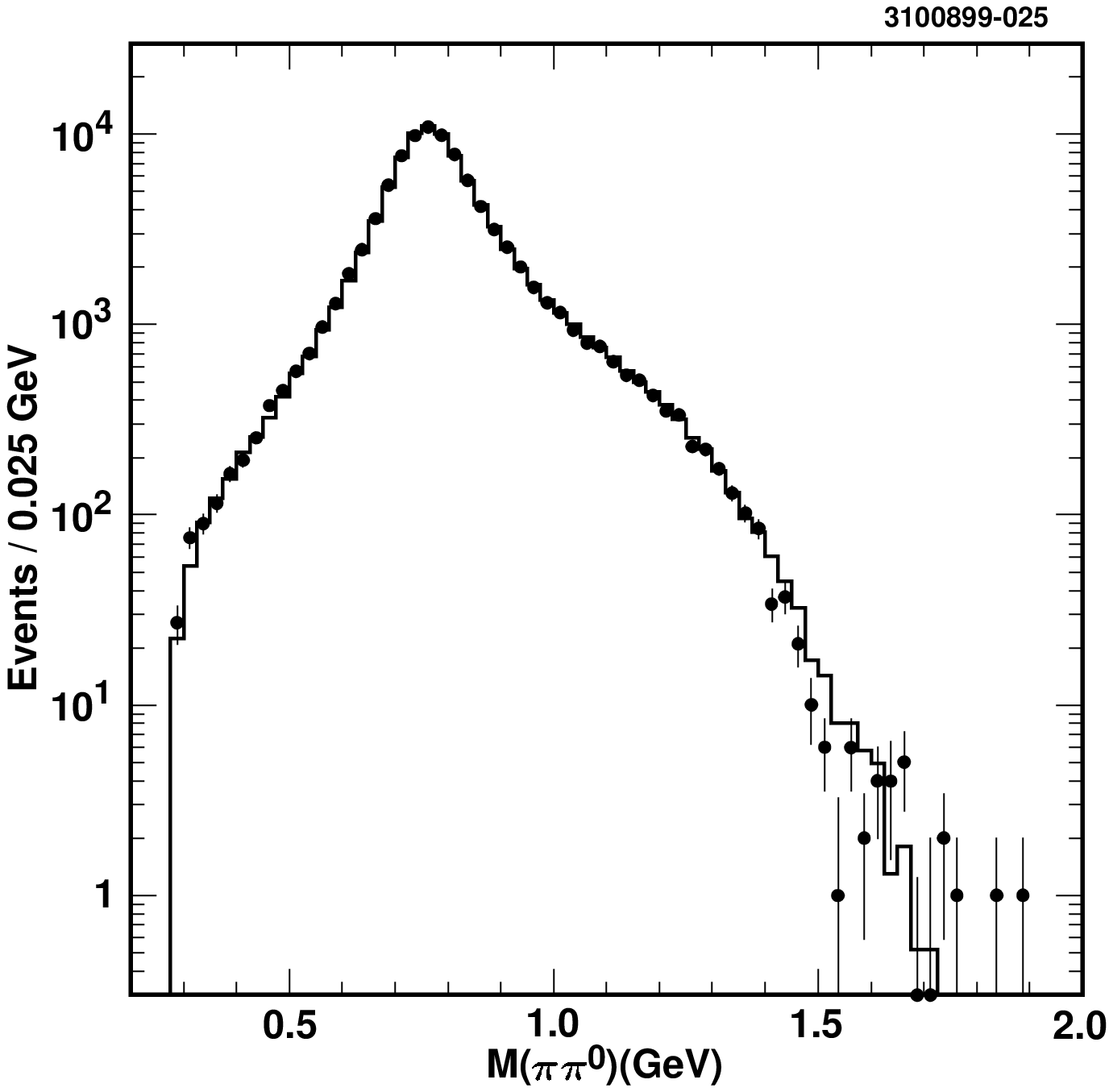}
   \caption{Raw $M_{\pi^-\pi^0}$ spectrum after $\pi^0$ side-band subtraction
            for candidate decays from the data (points) 
            and the generic $\tau$ Monte Carlo sample (line histogram).   
           }
  \label{fig:mpipi}
\end{figure}

\begin{figure}
   \leavevmode\centering 
   \epsfxsize=3.5in
   \epsfbox{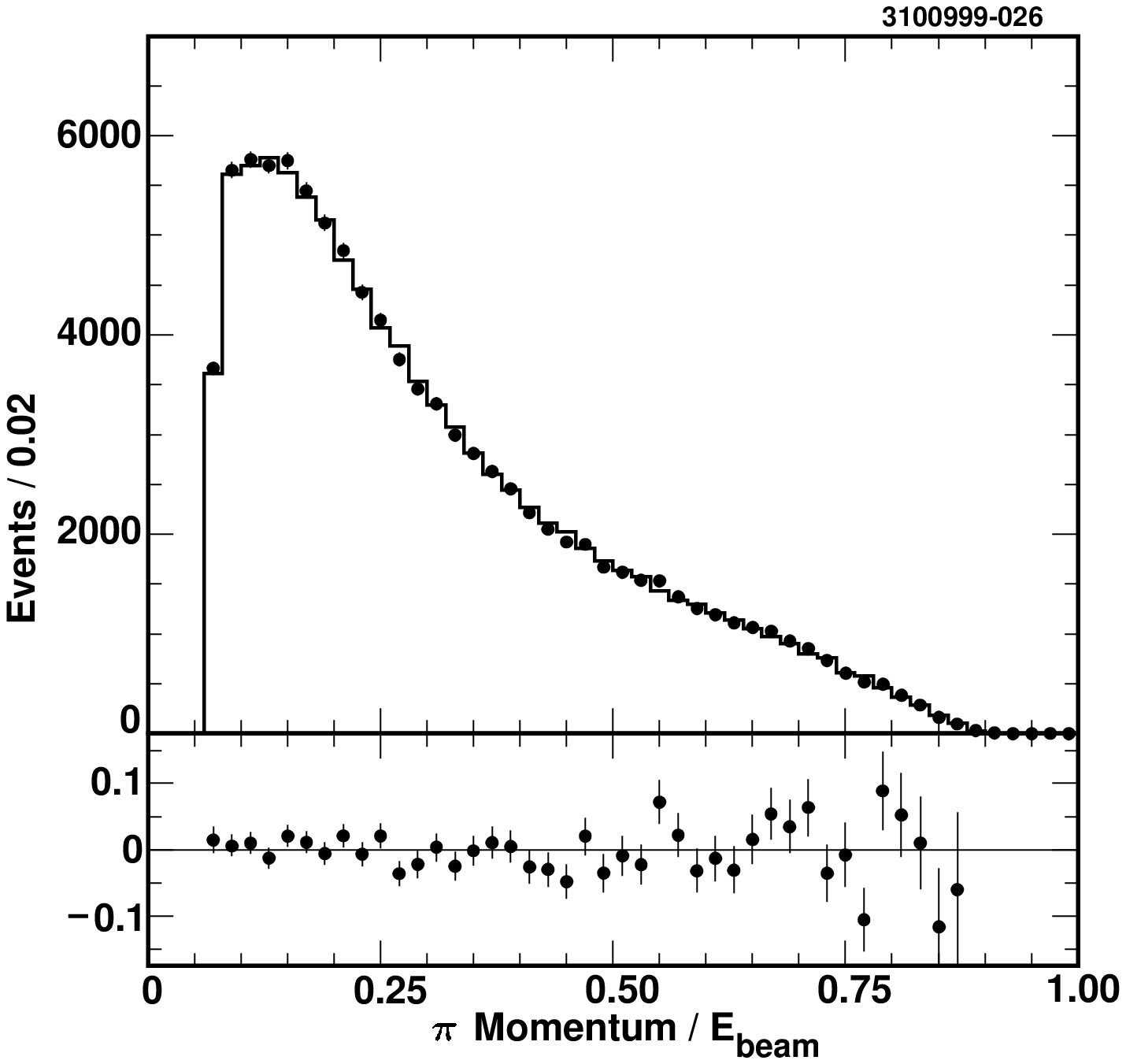}
   \caption{Distribution in the momentum of the $\pi^-$, divided by the 
            beam energy, 
            for candidate decays, from the data (points) 
            and the generic $\tau$ Monte Carlo (line histogram) samples, 
            after $\pi^0$ side-band subtraction.   The bottom plot 
            gives the deviations of the data spectrum from the Monte Carlo 
            spectrum, normalized by the Monte Carlo spectrum.  
           }
  \label{fig:ppi}
\end{figure}

\begin{figure}
   \leavevmode\centering 
   \epsfxsize=3.5in
   \epsfbox{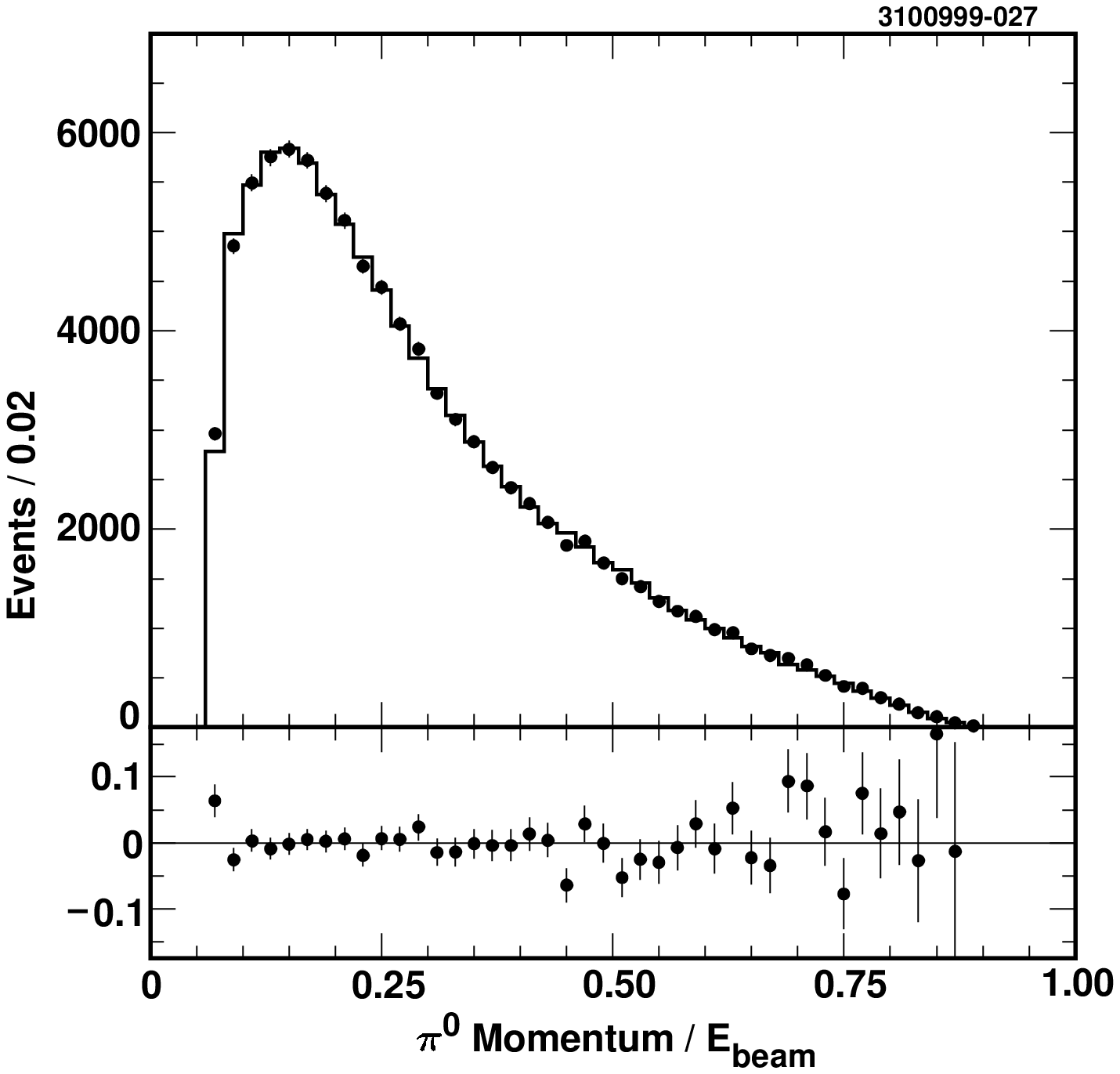}
   \caption{Distribution in the energy of the $\pi^0$, divided by the 
            beam energy, 
            for candidate decays, from the data (points) 
            and the generic $\tau$ Monte Carlo (line histogram) samples, 
            after $\pi^0$ side-band subtraction.   The bottom plot 
            gives the deviations of the data spectrum from the Monte Carlo 
            spectrum, normalized by the Monte Carlo spectrum.  
           }
  \label{fig:epi0}
\end{figure}

The $M_{\pi^-\pi^0}$ spectrum is shown, after side-band subtraction, 
in Fig.~\ref{fig:mpipi}.
The agreement between data and MC spectra is more than 
an indication of the validity of the application of CVC.  
It also suggests that the event kinematics in the MC samples
are sufficiently similar to those in the data that the selection 
criteria described above are not likely to have introduced significant 
biases.  Additional support for this is the comparison between data 
and generic Monte Carlo samples of the $\pi^-$ 
momentum and $\pi^0$ energy distributions, shown in Figs.~\ref{fig:ppi} 
and~\ref{fig:epi0}.  
Several events in Fig.~\ref{fig:mpipi} lie above the $\tau$ 
lepton mass.  
The small number of these events indicates that possible backgrounds 
at high mass, such as low-multiplicity $q\overline{q}$ events,  
are not significant. 

After $\pi^0$ side-band subtraction, 
backgrounds from non-signal $\tau$ decays are estimated to be 
$6.63\pm 0.20\,\%$ from the generic $\tau$ Monte Carlo sample.  
The dominant channels are 
$\tau^-\to \pi^-\pi^0\pi^0\nu_\tau$ ($4.01\pm 0.08\,\%$), 
$\tau^-\to K^-\pi^0\nu_\tau$ ($1.86\pm 0.16\,\%$) and 
$\tau^-\to h^-K^0_L\pi^0\nu_\tau$ ($0.59\pm 0.08\,\%$), 
where $h$ denotes $\pi$ or $K$ and the errors include 
branching fraction uncertainties as well as statistical 
errors.  


\section{CORRECTIONS TO THE DATA}
\label{s-corr}

One goal of this analysis is to analyze the $\pi^-\pi^0$ mass spectrum  
in the context of several models.  
However, we are not able to explore all possible models.  
In addition, 
it is desirable to compare our spectrum to data from other experiments 
in a model independent way, as well as to present it in a form that 
facilitates comparison with future data or models.  
These considerations motivate us to construct 
a histogram of the mass spectrum 
that has been {\sl corrected} for (primarily) experimental 
effects.  We then carry out simple $\chi^2$ fits to the corrected spectrum 
using the models of the $\pi^-\pi^0$ line shape described earlier.  

Three experimental effects give rise to distortions in the $\pi^-\pi^0$ 
mass spectrum: (1) backgrounds; (2) smearing due to
resolution and radiative effects; and (3) mass-dependence of the 
experimental acceptance.  In this section, we describe the corrections, 
that we applied in the order listed to remove these distortions.
These corrections rely on the Monte Carlo simulation of the physics and 
detector response.  

\subsection{Binning of the {\boldmath $M_{\pi^-\pi^0}$} Spectrum}

Before discussing the corrections mentioned above, we note that we have 
elected to bin the $M_{\pi^-\pi^0}$ spectrum in intervals of 25~MeV below 
1~GeV, and 50~MeV above 1~GeV.  This binning is chosen so as  
to be sensitive to rapidly varying regions of the spectrum while 
limiting the size of the bin migration correction and consequently 
the magnitude of correlations among nearby bins in the corrected spectrum.  
This is important for the stability and accuracy of the $\chi^2$ fit 
procedure, which is known to be biased when data points are strongly 
correlated~\cite{chibias}.  
The increase in mass resolution from approximately 6~MeV at low masses 
to 17~MeV at high masses motivates the large bin width above 1~GeV.  
The large bin width is also beneficial in the very high mass bins where 
low statistics could lead to non-Gaussian fluctuations.

\subsection{Corrections for Backgrounds}

\begin{figure}
   \leavevmode\centering 
   \epsfxsize=6.5in
   \epsfbox{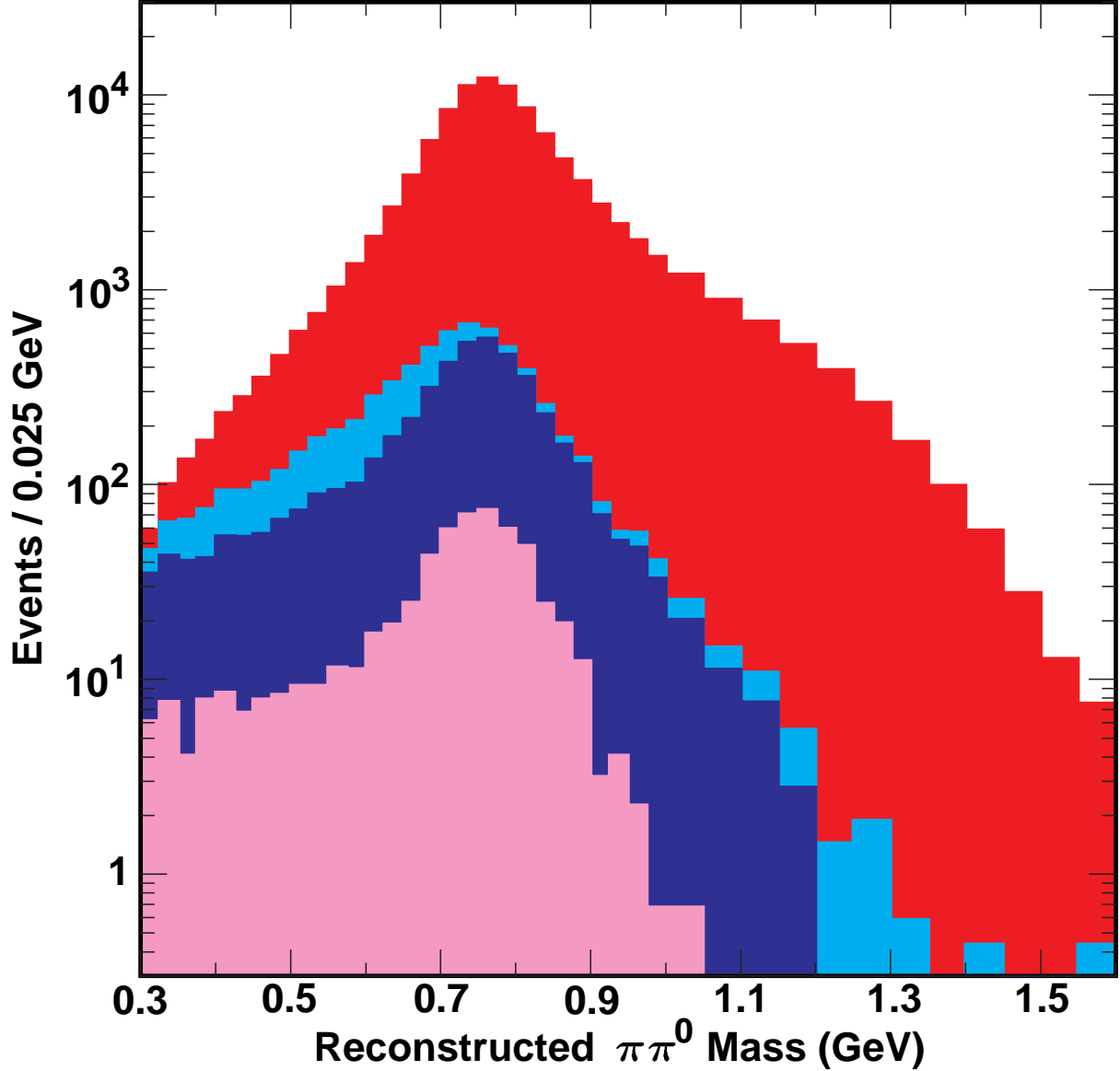}
   \caption{Contributions to the 
            reconstructed $M_{\pi^-\pi^0}$ spectrum from Monte Carlo 
            simulation of signal and background $\tau$ decays. From 
            bottom to top spectra for the following channels are plotted
            cumulatively: 
            $\tau\to K_L h\pi^0\nu$ ($h$ denoting $\pi$ or $K$),  
            $\tau\to \pi\pi^0\pi^0\nu$, $\tau\to K\pi^0 \nu$ (backgrounds), 
            and $\tau\to \pi\pi^0\nu$ (signal). 
           }
  \label{fig:bg}
\end{figure}

As noted earlier, the backgrounds entering the 
$\tau^-\to\pi^-\pi^0\nu_\tau$ sample are small.  Side bands in the 
$M_{\gamma\gamma}$ distribution are used to model the fake-$\pi^0$
contribution.  The remaining $6.6\%$ are modeled with the generic 
$\tau$-pair Monte Carlo sample.  These subtractions are performed 
bin by bin in the mass spectrum.  The Monte Carlo spectra for the 
signal and primary background modes are plotted in Fig.~\ref{fig:bg}, 
after $\pi^0$ side-band subtraction.

\subsection{Correction for Bin Migration}

Detector resolution causes the $\pi^-\pi^0$ mass spectrum to become
broader.   The presence of radiation in the decay 
$\tau^-\to\pi^-\pi^0(\gamma)\nu_\tau$ is also important.  The 
radiative photons tend to be low in energy, and are difficult to 
distinguish from photons from a possible second $\pi^0$ or from 
fragments of the hadronic shower from charged pions that 
interact in the calorimeter.
Consequently, we can not identify them reliably, either for inclusion 
in the invariant mass calculation or as a basis for vetoing events.  
The net effect of ignoring decay radiation is to 
broaden and shift the mass spectrum.  
 
\begin{figure}
   \leavevmode\centering 
   \epsfxsize=6.5in
   \epsfbox{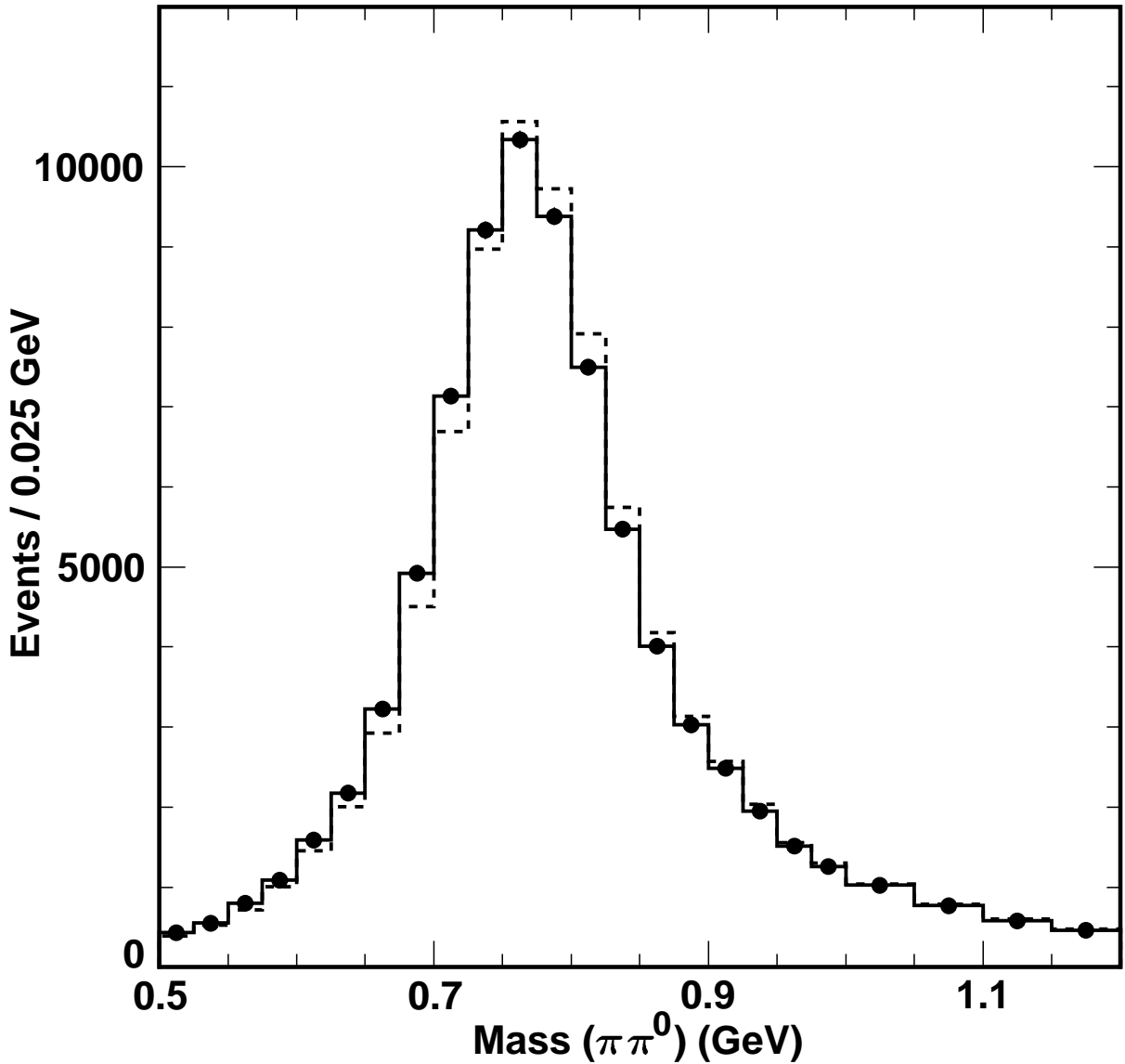}
   \caption{Effect of bin migration correction: $M_{\pi^-\pi^0}$ spectrum 
            prior to (solid histogram with points), and after 
            (dashed histogram) unfolding of resolution and radiative  
            distortions.
           }
  \label{fig:corr}
\end{figure}

We correct for these effects by performing an approximate unfolding 
procedure, based on a bin migration matrix determined from the 
$\tau^-\to\pi^-\pi^0\nu_\tau$ MC sample.  This procedure is outlined 
in Appendix~\ref{a-unfold}.  
Since the experimental resolution on $M_{\pi\pi^0}$ is dominated 
by that on the $\pi^0$ energy, the agreement between data and MC 
shown in Fig.~\ref{fig:spi0} gives us confidence in this aspect of the 
correction procedure.  For the radiative effect, we rely on the 
{\tt PHOTOS}-based simulation \cite{photos} employed by {\tt TAUOLA}.  
The unfolded spectrum is shown as the dashed histogram in 
Fig.~\ref{fig:corr}, with the uncorrected spectrum overlaid.  

\subsection{Correction for Acceptance}

\begin{figure}
   \leavevmode\centering 
   \epsfxsize=5.5in
   \epsfbox{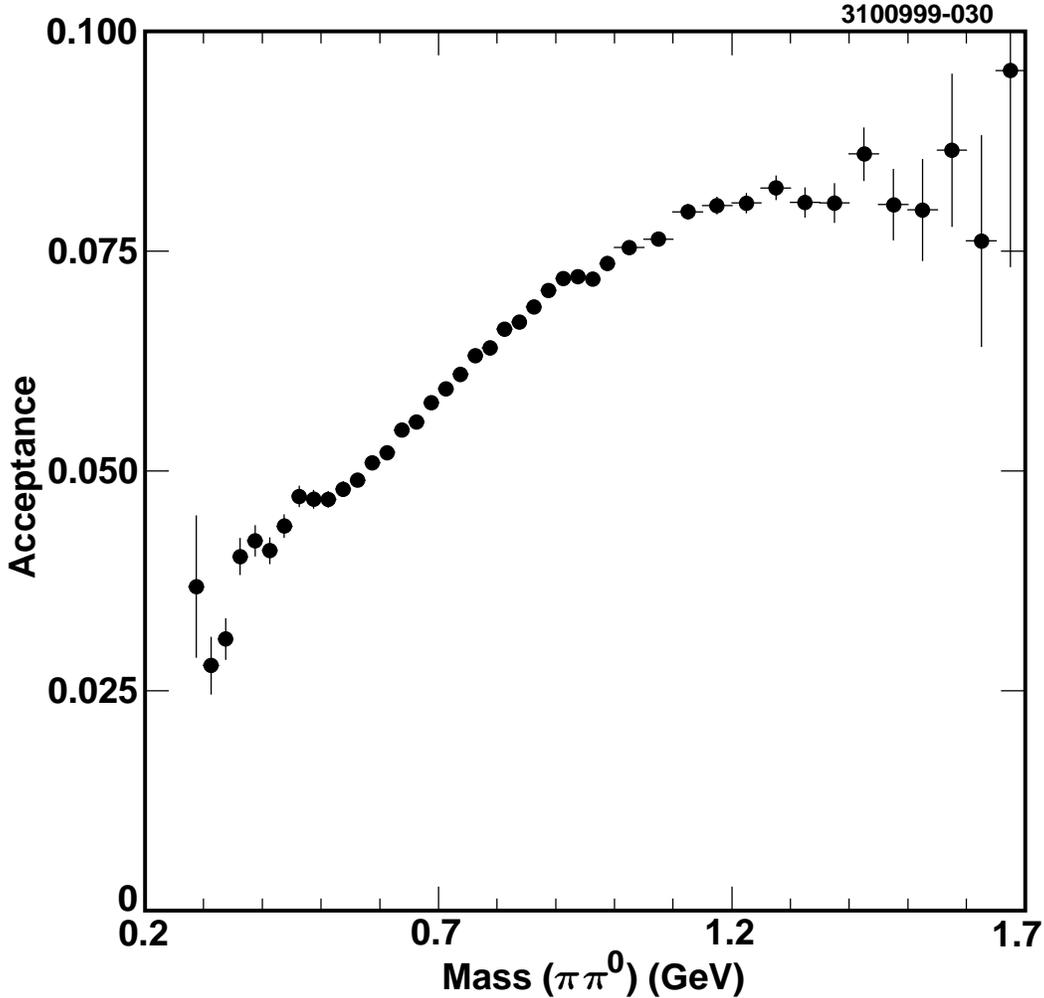}
   \caption{Acceptance as a function of generated $\pi^-\pi^0 (\gamma)$ 
            mass, as determined from the full $\tau^-\to\pi^-\pi^0\nu_\tau$ 
            Monte Carlo sample.  
           }
  \label{fig:acc}
\end{figure}

Finally we correct for mass dependence of the acceptance, plotted  
in Fig.~\ref{fig:acc}.  
Again, this is determined from the MC simulation.  The main effects 
causing this dependence are associated with the kinematics of the 
decay and the cuts imposed in the event selection, both of which are 
well understood.  
%

\subsection{The Corrected Mass Spectrum}

The corrected mass spectrum is given in tabular form in 
Appendix~\ref{a-massspectrum}, along with elements of the covariance 
matrix characterizing the statistical errors and the correlations among 
entries introduced by the bin migration correction procedure.  The spectrum 
is also made available electronically~\cite{wwwrho}.  

%
%
%

\section{RESULTS OF FITS FOR RESONANCE PARAMETERS}
\label{s-results}

\subsection{Fitting Procedure}

We perform $\chi^2$ fits to the fully-corrected 
$\pi^-\pi^0$ mass spectrum to extract resonance parameters and 
couplings.  The $\chi^2$ minimization and parameter error 
determination is carried out using the {\tt MINUIT} 
program~\cite{minuit}.  
Because of
poor statistics and/or uncertainties associated with the background
estimation and acceptance correction, 
only data in the range 0.5 to 1.5 GeV are included
in the fits.  Also, as a result of the unfolding procedure, 
off-diagonal terms of the covariance matrix are non-zero, and the
corresponding terms must be included in the calculation of 
the $\chi^2$~\cite{dsinv}.  
The off-diagonal elements of the covariance matrix are 
not reflected in the error bars shown in the figures in this section. 

Since the 
functional forms used to fit the data are nonlinear functions of many 
parameters, several iterations of minimization are performed before 
convergence is reached.  We also integrate the fit function within each 
bin when computing the $\chi^2$.  We have tested this procedure using 
high-statistics generator-level Monte Carlo samples to ensure the 
reproducibility and accuracy in the determination of fit parameters and 
their errors.

\subsection{Fit to the K\&S Model}
\label{ss-ksfit}

In this section, we report in detail the results from the simplest fit,  
done using the K\&S model with no $\rho^{\prime\prime}$ contribution.  
Although this model is normalized according to the $F_\pi(0)=1$ constraint, 
we introduce an additional parameter multiplying the K\&S function 
(Eq.~\ref{eq:vks}), which is allowed to float in the fit.  We have elected 
to do this for several reasons.  First, we do not expect this or any other 
model of the line shape of a broad resonance to hold arbitrarily far from 
its peak.  If the model does not hold at very low values of 
$M_{\pi^-\pi^0}$ then enforcing the $F_\pi(0)=1$ condition can bias the 
resonance parameters.  Second, the focus of this analysis has been on the 
shape of the mass spectrum: for example, tight cuts have been applied to 
maintain high purity.  The normalization of this spectrum 
(see Sec.~\ref{ss-compind}) depends on external measurements which have 
experimental uncertainties, as well as on theoretical factors which also 
have uncertainties.  Given these considerations, the value of 
strictly enforcing the normalization condition is questionable. 

\begin{figure}
   \leavevmode\centering 
   \epsfxsize=5.5in
   \epsfbox{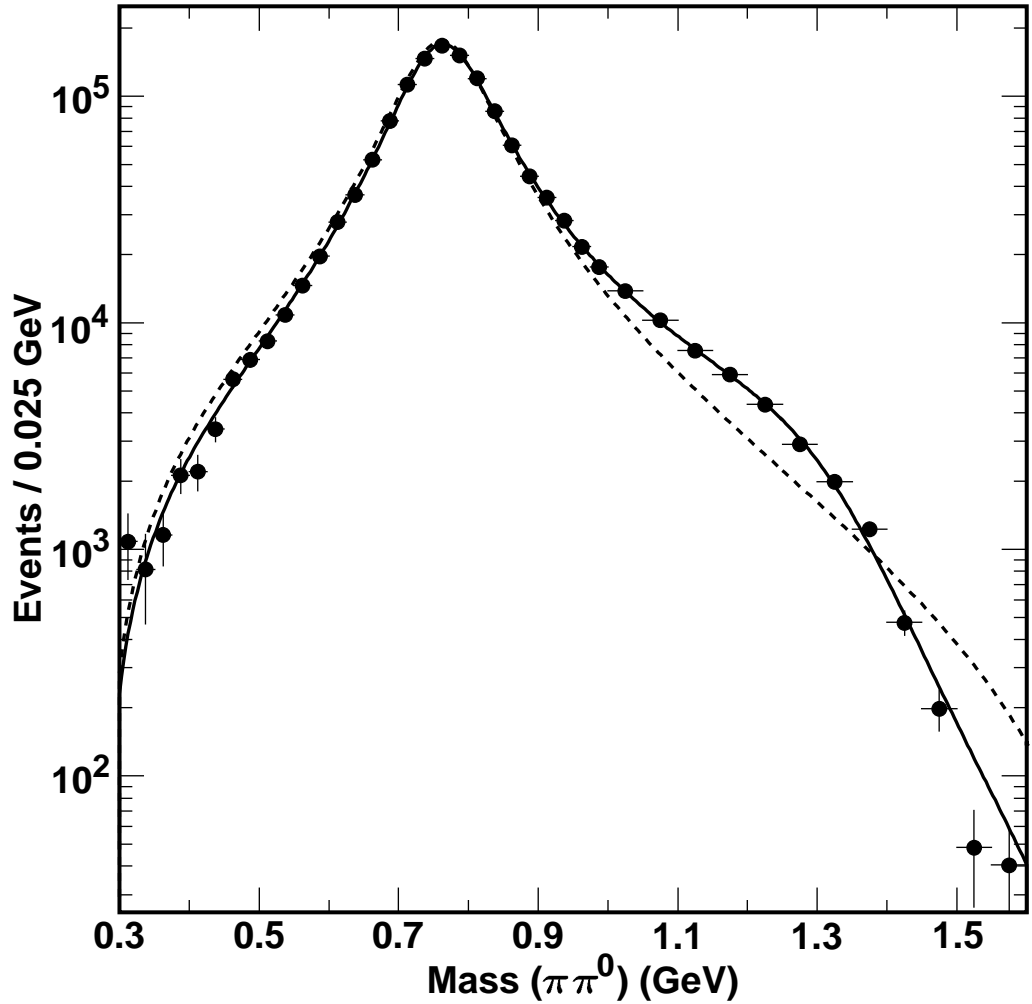}
   \caption{Fully corrected $M_{\pi^-\pi^0}$ distribution in 
            $\tau^-\to\pi^-\pi^0\nu_\tau$ events (points).
            The solid curve overlaid represents the results of 
            the fit to the K\&S model.  The dashed curve is obtained
            using the $\rho(770)$ parameters obtained from this fit, 
            but with the $\rho^\prime$ contribution turned off
            ({\sl i.e.}, $\beta$ set to zero).
           }
  \label{fig:finalfit}
\end{figure}

The fully corrected $M_{\pi\pi^0}$  
spectrum with the fit function superimposed is displayed 
in Fig.~\ref{fig:finalfit}.
The $\chi^2$ for this fit is 27.0 for 24 degrees of freedom.  
We obtain
\begin{eqnarray*}
        M_\rho               & = &  774.9 \pm 0.5 \pm 0.9 \; {\rm MeV}, \\
        \Gamma_\rho          & = &  149.0 \pm 1.1 \pm 0.7 \; {\rm MeV}, \\
        \beta                & = & -0.108 \pm 0.007 \pm 0.005, \\
        M_{\rho^\prime}      & = & 1364   \pm  7  \pm  8 \; {\rm MeV}, \\
        \Gamma_{\rho^\prime} & = &  400   \pm 26  \pm 23 \; {\rm MeV}, 
\end{eqnarray*}
where the first error is statistical and the second is due to 
systematic uncertainties, described in Sec.~\ref{s-syserr}.  When 
interpreted in terms of the pion form factor, the normalization gives 
$|F_\pi(0)|^2 = 1.16\pm 0.02$, where the error is statistical only.  
Performing the same fit, but with the $F_\pi(0) = 1$ normalization condition 
imposed, yields a $\chi^2$ of 62.1 for 25 degrees of freedom and significantly 
different values for the other fit parameters 
({\sl i.e.}, $M_\rho = 772.3\,\mbox{\rm MeV},\;
\Gamma_\rho = 144.6\,\mbox{\rm MeV}$).

The fit parameters are correlated, with the correlation matrix:
\begin{equation}
  \left(\begin{array}{rrrrrr}
        1.00  &        &        &        &       &       \\
        0.64  &  1.00  &        &        &       &       \\
        0.75  &  0.51  &  1.00  &        &       &       \\
        0.89  &  0.68  &  0.43  &  1.00  &       &       \\
        0.21  &  0.22  &  0.21  &  0.16  & 1.00  &       \\
       -0.52  & -0.38  & -0.08  & -0.74  
                           &  \phantom{-}0.33  & \phantom{-}1.00 
   \end{array}\right), 
\end{equation}
where the parameters are normalization, $M_\rho$, $\Gamma_\rho$, 
$\beta$, $M_{\rho^\prime}$, and $\Gamma_{\rho^\prime}$, respectively.

\subsection{Fits to Other Models}

\begin{figure}
   \leavevmode\centering 
   \epsfxsize=5.5in
   \epsfbox{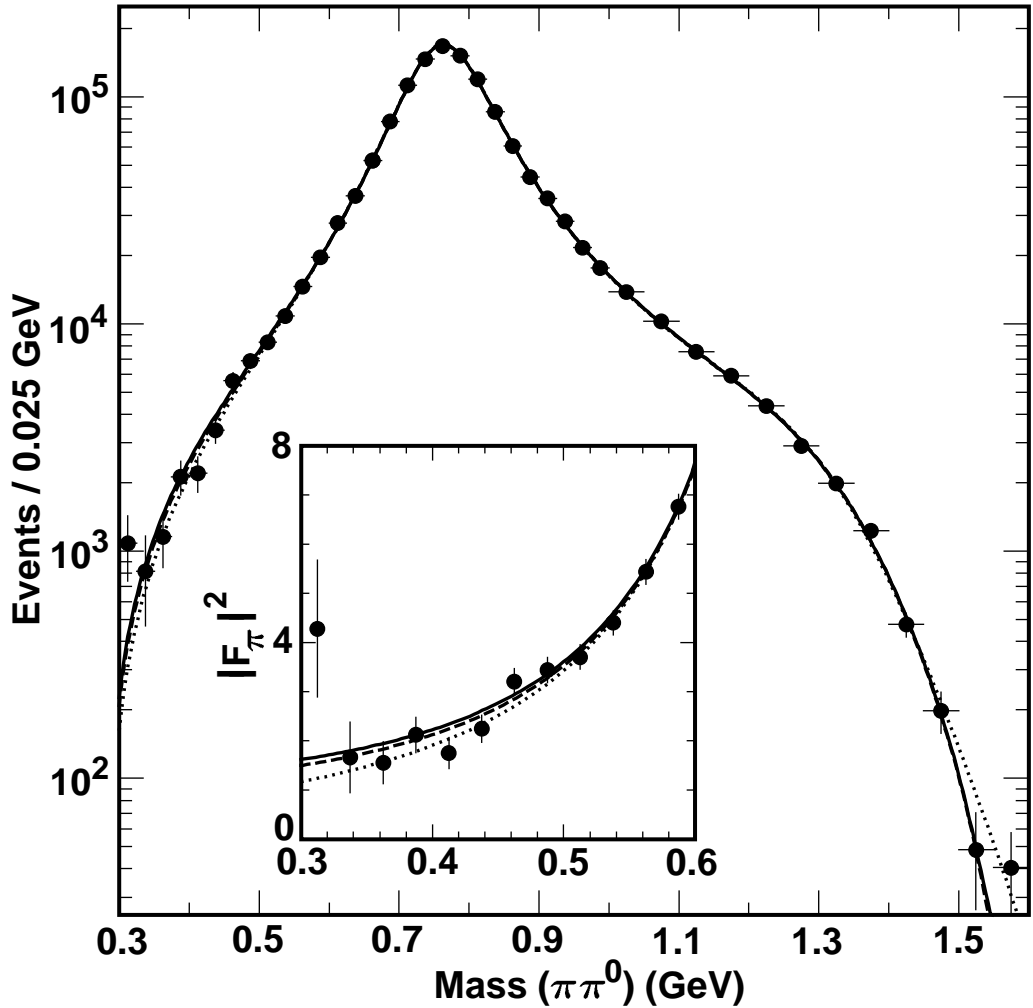}
   \caption{Alternate fits to the fully corrected $M_{\pi^-\pi^0}$ 
            distribution in $\tau^-\to\pi^-\pi^0\nu_\tau$ events (points).
            The solid curve overlaid represents the results of 
            the fit to the K\&S model with $\rho^{\prime\prime}$ contribution 
            included.  The dashed curve represents the fit to the G\&S model, 
            also with the $\rho^{\prime\prime}$.
            The dotted curve is the fit to the 
            K\&S model, including the Blatt-Weisskopf barrier factor but 
            no $\rho^{\prime\prime}$.  The inset shows the low mass region 
            where the differences between the models are most significant. 
            Here we plot $|F_\pi|^2$ (see Sec.~\ref{ss-compind}), 
            eliminating the purely kinematic factors which cause 
            rapid variation in $M_{\pi^-\pi^0}$ in this region. 
           }
  \label{fig:fits}
\end{figure}

The results from fits of the corrected $M_{\pi^-\pi^0}$ spectrum to various 
models are given in Table~\ref{tab:results}.  
\begin{table}
\caption[]{Results from fits to the corrected $M_{\pi^-\pi^0}$ spectrum 
           over the range 0.5--1.5~GeV, for several models.  The errors 
           shown are statistical only.  K\&S refers to the model of 
           K\"uhn and Santamaria~\cite{KS} while G\&S refers to the model 
           of Gounaris and Sakurai~\cite{GS}.  See text for descriptions 
           of these models and their variants.
          }
\label{tab:results}
\begin{tabular}{lccccc}
    Fit Parameter & \multicolumn{5}{c}{Model} \\
                  &  K\&S  & K\&S w/$\rho^{\prime\prime}$ 
                  &  K\&S w/barrier & G\&S & G\&S w/$\rho^{\prime\prime}$ \\
\hline
  $M_\rho$ (MeV)  
                  & $774.9\pm 0.5$ & $774.6\pm 0.6$  
                  & $769.7\pm 0.7$ & $775.3\pm 0.5$ & $775.1\pm 0.6$ \\
  $\Gamma_\rho$ (MeV) 
                  & $149.0\pm 1.1$ & $149.0\pm 1.2$ 
                  & $145.8\pm 1.3$ & $150.5\pm 1.1$ & $150.4\pm 1.2$ \\
  $\beta$
                  & $-0.108\pm 0.007$ & $-0.167\pm 0.008$ 
                  & $-0.160\pm 0.008$ & $-0.084\pm 0.006$ & $-0.121\pm 0.009$ \\
  $M_{\rho^\prime}$ (MeV) 
                  & $1364\pm 7$    & $1408\pm 12$ 
                  & $1321\pm 9$    & $1365\pm 7$    & $1406\pm 13$ \\
  $\Gamma_{\rho^\prime}$ (MeV)   
                  & $400\pm 26$    & $502\pm 32$
                  & $397\pm 17$    & $356\pm 26$    & $455\pm 34$ \\
  $R_\rho$ (GeV$^{-1}$)
                  & ---            & ---
                  & $1.9 \pm 0.3$  & ---            & --- \\
  $R_{\rho^\prime}$ (GeV$^{-1}$)
                  & ---            & ---
                  & $5.0 \pm 2.2$  & ---            & --- \\
  $\gamma$
                  & $\equiv 0$     & $0.050\pm 0.010$
                  & $\equiv 0$     & $\equiv 0$     & $0.032\pm 0.009$ \\
  $ | F_\pi(0) |^2$
                  & $1.16\pm 0.02$ & $1.14\pm 0.02$ 
                  & $0.39\pm 0.01$ & $1.04\pm 0.02$ & $1.03\pm 0.02$ \\
\hline
 $\chi^2$/dof 
                  & 27.0/24        & 23.2/23
                  & 22.9/22        & 26.8/24        & 22.9/23 \\
\end{tabular}
\end{table}
%
Several of these fits are illustrated in Fig.~\ref{fig:fits}.
For fits including a $\rho^{\prime\prime}$ contribution, we fix its parameters 
to world average values~\cite{pdg98}
($M_{\rho^{\prime\prime}} = 1.700\,\mbox{\rm GeV},\; 
\Gamma_{\rho^{\prime\prime}} = 0.235\,\mbox{\rm GeV}$),
but allow the relative coupling constant $\gamma$ to float.  

The G\&S fits behave similarly to the K\&S fits, as suggested by the 
$\chi^2$ values in Table~\ref{tab:results}, and by the nearly overlapping 
solid and dashed curves in Fig.~\ref{fig:fits}.  In the figure, the 
deviation between the K\&S and G\&S curves is only visible at very low 
and very high values of $M_{\pi^-\pi^0}$.  The deviation at low values 
is reflected by the difference in the inferred extrapolations of $|F_\pi|^2$ 
to $q^2 = 0$, where the G\&S model gives results more consistent with the 
expectation $F_\pi(0) = 1$.  

The presence of the $\rho^{\prime\prime}$ in $e^+e^-\to\pi^+\pi^-$ 
is evident from the cross section measurements of 
DM2~\cite{dm2} near and above $M_\tau$.  For $\tau$ lepton decay, 
the $\rho^{\prime\prime}$ pole mass is near the endpoint of the 
$M_{\pi^-\pi^0}$ spectrum, thus making it difficult to observe.  
However, as with the $\rho^{\prime}$, its influence can be observed 
as an interference effect.  While we obtain good fits without the 
$\rho^{\prime\prime}$ meson, the $\chi^2$ values for both K\&S and 
G\&S models are significantly improved when such a contribution is introduced.  
It can be seen in Fig.~\ref{fig:fits} that the $\rho^{\prime\prime}$ fits 
also agree better with the data points in the 1.5--1.6 GeV region which were 
excluded from the fits.  

The relative phases of $\rho$, $\rho^\prime$ and $\rho^{\prime\prime}$ which 
we have obtained as $(+\,-\,+)$ are consistent with expectations from some 
models~\cite{CD,CIL}.  We also find that including the $\rho^{\prime\prime}$ 
has a significant impact on our measurement of the $\rho^\prime$ parameters.  
In particular, values for the $\rho^\prime$ mass are closer to those 
based on other decay modes~\cite{pdg98,smu} when the 
$\rho^{\prime\prime}$ is included.  

We have also modified the energy dependence of the $\rho$ and $\rho^{\prime}$ 
widths by including the Blatt-Weisskopf factor (see Eq.~\ref{eq:blatt}).  
The results for the nominal K\&S fit function 
(with no $\rho^{\prime\prime}$ contribution)
modified in this way are given in 
Table~\ref{tab:results} and shown as the dotted curve in Fig.~\ref{fig:fits}.
Values obtained for the $\rho$ and $\rho^{\prime}$ range parameters 
$R_\rho$ and $R_{\rho^\prime}$ are consistent with expectations.  
This fit yields a smaller $\chi^2$ per degree of freedom than other 
fits that also do not include the $\rho(1700)$.  
However, this function as implemented does not yield 
a normalization consistent with $F_\pi(0)=1$.


\section{SYSTEMATIC ERRORS}
\label{s-syserr}

Systematic errors are listed in Table~\ref{tab:syserr}.  These have been 
determined for the nominal K\&S fit, however they are representative of 
those associated with G\&S type fits as well.  
We have not included uncertainties associated with model dependence
in our systematic error assessment.  
We discuss the most significant sources of error below.
\begin{table}[hbt]
\caption{Systematic errors, in MeV except those for $\beta$, which is 
         dimensionless.
        }
\label{tab:syserr}
  \begin{tabular}{lrrrrr}
      \multicolumn{1}{c}{Source} 
    & \multicolumn{1}{c}{$M_\rho$ }
    & \multicolumn{1}{c}{$\Gamma_{\!\rho}$ }
    & \multicolumn{1}{c}{$\beta$ }
    & \multicolumn{1}{c}{$M_{\rho^\prime}$ }
    & \multicolumn{1}{c}{$\Gamma_{\!\rho^\prime}$ } \\
\hline
  & & & & &\\
   $\!\!$Backgrounds       & 0.1 & 0.4 & 0.002    &  1 &  5 \\
   $\!\!$Bin Migration     & 0.3 & 0.5 & 0.001    &  5 & 16 \\
   $\!\!$Momentum Scale    & 0.2 & 0.1 & $<0.001$ &  1 &  1 \\
   $\!\!$Energy Scale      & 0.8 & 0.2 & $0.002$  &  2 &  6 \\
   $\!\!$Acceptance        & 0.2 & 0.2 & 0.004    &  5 & 15 \\
   $\!\!$Fit Procedure     & 0.1 & 0.2 & $<0.001$ &  1 &  1 \\
\hline
  & & & & &\\
   $\!\!$Total Syst.       & 0.9 & 0.7 & 0.005    &  8 & 23 \\
   $\!\!$Stat.~Error       & 0.5 & 1.1 & 0.007    &  7 & 26 \\
  \end{tabular}
\end{table}

\subsection{Background Subtraction}

We rescale the individual background components by amounts 
consistent with uncertainties in branching fractions and detection 
efficiency.  Typically, this is $\pm (5\mbox{\rm --}10)\%$ of the nominal.  
The largest change in fit parameters comes about by varying the 
$\tau^-\to K^-\pi^0\nu_\tau$ contribution in this way.  

\subsection{Bin Migration Correction}

The resolution on $M_{\pi^-\pi^0}$ 
is dominated by the photon energy
measurement.  The excellent agreement between data and MC
seen in the $\pi^0$ mass spectra in Figure~\ref{fig:spi0} gives us
confidence that this resolution function is adequately simulated.  
That the MC correctly models radiative effects is tested by comparing
characteristics of photon-like showers accompanying the $\pi^-\pi^0$ 
system with those in the data.  

The method applied to unfold these effects relies on the 
approximate similarity between the $M_{\pi^-\pi^0}$ line shape used 
as input to the MC and the true one (see Appendix~\ref{a-unfold}).  
We have estimated the bias 
resulting from the inaccuracy of this approximation, and find it 
to be small.  Considering this bias and possible errors in the 
modeling of the effects themselves, we conservatively arrive at 
the uncertainties shown in Table~\ref{tab:syserr}.  For reference, 
failure to correct for bin migration results in values
for $M_\rho$ and $\Gamma_{\!\rho}$ which are 2.6 MeV lower and 4.6 MeV 
higher, respectively.  Thus we believe we understand this correction to 
$\sim 10\%$ of itself.

\subsection{Energy and Momentum Scales}

Although the $\pi^0$ mass resolution function is well modeled, 
even a small error in the 
absolute energy calibration of the calorimeter 
can have an effect on $M_\rho$ 
without grossly disturbing the agreement in Figure~\ref{fig:spi0}.  
Based on studies of photons from various processes, we believe this 
calibration to be good to better than 0.3\%~\cite{ajwmass}.  
Scaling the photon energies in the data by a factor of $1\pm 0.003$ 
and fitting
the $M_{\pi^-\pi^0}$ spectra thus obtained results in the 
uncertainties shown in Table~\ref{tab:syserr}.  

The same procedure 
is used to assess the systematic error due to the absolute momentum
scale uncertainty.   Studies of $D$ meson decays and 
$e^+e^-\to \mu^+\mu^-$ events have demonstrated that the 
momentum scale uncertainty is below $0.05\%$~\cite{bkhmev}.

\subsection{Other Sources of Error}

To estimate errors associated with the mass dependence of the detection 
efficiency, we modified the shape of the acceptance correction distribution 
(Fig.~\ref{fig:acc}) by amounts suggested by the variation of selection 
criteria.  We also performed the full analysis after varying cuts on 
$S_{\gamma\gamma}$, minimum photon energy, and the $\pi^0\to \gamma\gamma$ 
decay angle.  Uncertainties associated with the fitting procedure were 
evaluated by performing fits to generator-level Monte Carlo spectra (no 
detector effects simulated).  We also fit the uncorrected data and MC 
spectra, using the observed shifts in fit parameters for the MC sample to 
correct the parameters obtained from the data sample.  This procedure 
yielded results that were in agreement with our nominal procedure.  
As a final cross-check, we split the data sample according to the tag 
decay.  The results obtained for the three tags were in agreement with 
each other.  

%

\section{COMPARISON WITH DATA FROM OTHER EXPERIMENTS}
\label{s-compexp}

\subsection{Comparison with Fits by Other Experiments}

The results from our fits can be compared with similar fits 
by ALEPH~\cite{aleph} to their data, as well as with fits by 
other authors~\cite{KS,ben,barkov} to various compilations of 
$e^+e^-\to\pi^+\pi^-$ data.  For illustration purposes, 
some representative comparisons are given in Table~\ref{tab:compfit}.  
\begin{table}
\caption[]{Comparison of results of fits using the K\&S and G\&S models 
           ($\rho^{\prime\prime}$ included with 
           $M_{\rho^{\prime\prime}},\ \Gamma_{\rho^{\prime\prime}} = 
           1700,\ 235$~MeV) to (a) CLEO $\tau$ data, (b) ALEPH $\tau$ 
           data~\cite{aleph} and (c) $e^+e^-\to \pi^+\pi^-$ data 
           (denoted `K\&S Fit', from fits in Ref.~\cite{KS}).  Units 
           are MeV for all fit parameters, except $\beta$, $\gamma$ and 
           $|F_\pi(0)|^2$ which are dimensionless.  Shown in parentheses 
           are the experimental uncertainties (statistical plus systematic). 
           The parameter  
           associated with $\rho$-$\omega$ interference in the $e^+e^-$ data 
           is not shown.
          }
\label{tab:compfit}
\begin{tabular}{lcccccc}
    Parameter     & \multicolumn{3}{c}{K\&S Model} 
                  & \multicolumn{3}{c}{G\&S Model} \\
                  & CLEO  & ALEPH & K\&S Fit 
                  & CLEO  & ALEPH & K\&S Fit \\
\hline
  $M_\rho$
                  & $774.6\,(1.1)$ & $774.9\,(0.9)$ & 773
                  & $775.1\,(1.1)$ & $776.4\,(0.9)$ & 776 \\
  $\Gamma_\rho$
                  & $149.0\,(1.4)$ & $144.2\,(1.5)$ & 144
                  & $150.4\,(1.4)$ & $150.5\,(1.6)$ & 151 \\
  $\beta$
                  & $-0.167\,(10)$ & $-0.094\,(7)$  & $-0.103$
                  & $-0.121\,(10)$ & $-0.077\,(8)$  & $-0.052$ \\
  $M_{\rho^\prime}$
                  & $1408\,(14)$   & $1363\,(15)$   & 1320
                  & $1406\,(15)$   & $1400\,(16)$   & 1330 \\
  $\Gamma_{\rho^\prime}$
                  & $502\,(39)$    & $\equiv 310$   & $390$   
                  & $455\,(41)$    & $\equiv 310$   & $270$ \\
  $\gamma$
                  & $0.050\,(10)$   & $-0.015\,(8)$  & $-0.037$
                  & $0.032\,(9)$   & $ 0.001\,(9)$  & $-0.031$ \\
  $ | F_\pi(0) |^2$
                  & $1.14\,(2)$    & $\equiv 1$     & $\equiv 1$
                  & $1.03\,(2)$    & $\equiv 1$     & $\equiv 1$ \\
\hline
 $\chi^2$/dof 
                  & 23.2/23        & 81/65          & 136/132
                  & 22.9/23        & 54/65          & 151/132 \\
\end{tabular}
\end{table}
For the $e^+e^-$ data we give the results of the 
fits carried out by K\"uhn and Santamaria~\cite{KS} to data below 
$\sqrt{s}\sim 1.6$~GeV.  These authors did not quote uncertainties 
on the parameters they obtained.  However, based on similar fits 
performed by 
Barkov~{\sl et al.}~\cite{barkov} and by us, we believe these 
uncertainties to be similar in magnitude to those from the $\tau$ decay 
data fits.  For example, averaging over models, Ref.~\cite{barkov} obtains 
$M_\rho = 775.9\pm 0.8\pm 0.8$~GeV, 
$\Gamma_\rho = 150.5\pm 1.6\pm 2.5$~GeV, 
where the first error represents that due to statistics plus systematics, 
and the second error gives the uncertainty due to model dependence.

The comparisons given in Table~\ref{tab:compfit} are not meant to be 
rigorous.  Due to the different assumptions made by different authors, 
a systematic comparison is not possible.  For example, the choice of 
whether to enforce the $F_\pi(0)=1$ condition for the K\&S model 
has a strong impact on the fit parameters and the goodness of fit 
for the $\tau$ data (see Sec.~\ref{ss-ksfit}).  
In the G\&S model this choice is less crucial 
since the values of $|F_\pi(0)|^2$ obtained when it is allowed to 
float are closer to unity than they are in the K\&S model.  
In addition, the fits to the $e^+e^-$ data shown do not include the 
high-$\sqrt{s}$ data from DM2~\cite{dm2}.  With this data included, 
fits we have carried out yield results that are considerably 
different from those with this data excluded.  The applicability 
of any given model across the full range of $\sqrt{s}$ accessible 
to experiments has not been demonstrated.  

Finally, small differences between charged and neutral $\rho$ meson 
parameters are expected, due to manifestations of isospin violation 
such as the $\pi^-/\pi^0$ mass difference.  
Refs.~\cite{adh} and~\cite{aleph} include discussions of this issue.  
Additional comments are given in Appendix~\ref{a-amu}.  

With these caveats, Table~\ref{tab:compfit} 
demonstrates general agreement between CLEO and ALEPH data, and between 
the $\tau$ and $e^+e^-$ data, supporting the applicability of CVC.  
In particular, 
the Gounaris-Sakurai fits show very good agreement between CLEO and ALEPH 
data.  Within the K\&S model, the smaller values of $\Gamma_\rho$  
obtained for the ALEPH and $e^+e^-$ data are likely a consequence of the 
$F_\pi(0)=1$ constraint, as similar values are also obtained when 
this constraint is imposed for the CLEO data (see Sec.~\ref{ss-ksfit}).

\subsection{Model Independent Comparisons}
\label{ss-compind}

\subsubsection{Comparison of CLEO and ALEPH data}

\begin{figure}
   \leavevmode
   \centering 
   \epsfxsize=4.0in
   \epsfbox{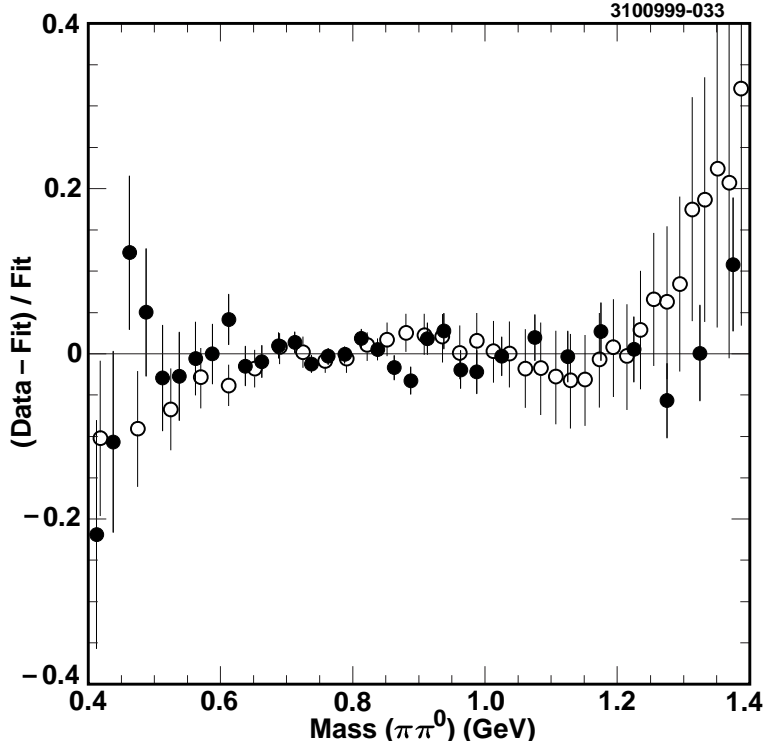}
   \caption{Difference in $\overline{v}(M_{\pi^-\pi^0})$ of 
            CLEO (filled circles) and ALEPH (open circles) data 
            from the fit of the CLEO 
            $\tau^-\to\pi^-\pi^0\nu_\tau$ data to the 
            Gounaris-Sakurai model, with $\rho^{\prime\prime}$ 
            included, divided by the fit value. 
           }
   \label{fig:cleo_aleph}
\end{figure}

It is also useful to compare data from different experiments in a 
model-independent way.  The ALEPH Collaboration has published their  
corrected $M^2_{\pi^-\pi^0}$ spectrum~\cite{aleph}, which may be 
compared directly with the corresponding CLEO spectrum.  Given the 
differences in binning, a visually useful comparison is obtained by 
determining the bin-by-bin deviations of the two data sets 
from the prediction of a given model.  For the prediction 
we use the results from the fit of the CLEO data to the 
G\&S model, including $\rho^{\prime\prime}$.  
In Fig.~\ref{fig:cleo_aleph}, we plot the deviations 
of the CLEO and ALEPH data from this model as a function 
of $M_{\pi^-\pi^0}$.  
The clustering of the CLEO points around zero is expected given 
the goodness of the fit.  The ALEPH points also cluster around 
zero, although systematic deviations may be present at low and high 
masses.  The significant correlations among the ALEPH points make 
this difficult to establish from the figure.  Independent of 
possible deviations from the model, however, the CLEO and ALEPH 
points show agreement over the entire spectrum.  

\subsubsection{Comparison of CLEO and $e^+e^-$ data}

Direct comparison, independent of model, of $\tau$ and $e^+e^-$
data can also be made as a further test of CVC.  To do this we 
represent the fully corrected CLEO $\pi^-\pi^0$ mass spectrum in terms 
of $|F_\pi|^2$ which can be compared with the $e^+e^-$ data directly. 
Using the leptonic $\tau$ decay width  
$\Gamma_e = G_F^2 M_\tau^5 S^{e}_{EW}/192\pi^3$ and 
the ratio of $\tau$ decay branching fractions to $\pi^-\pi^0\nu_\tau$  
($B_{\pi\pi^0}$) and $e^-\overline{\nu}_e\nu_\tau$ ($B_e$) for normalization, 
we derive the spectral function averaged over each mass bin 
from Eq.~\ref{eq:dgam}: 
\begin{equation}
  \overline{v}^{\pi\pi}(M_i) = \frac{B_{\pi\pi^0}}{B_e} \,
                    \frac{M_\tau^8}{12\pi\, |V_{ud}|^2}\,
                    \frac{S^e_{EW}}{S^{\pi\pi}_{EW}}\,
                    \frac{1}{M_i\,(M_\tau^2 - M_i^2)^2\,(M_\tau^2 + 2 M_i^2)}\,
                    \frac{1}{N}\,
                    \frac{N_i}{\Delta M_i}, 
\label{eq:specfun}
\end{equation}
where $M_i$ is the central value of $M_{\pi^-\pi^0}$ for the $i^{th}$ bin, 
and the quantity $1/N\,N_i/\Delta M_i$ is the number of entries ($N_i$) 
in the $i^{th}$ bin of the corrected measured $\pi^-\pi^0$ mass spectrum, 
divided by the total number of entries ($N$) and the bin width ($\Delta M_i$).
We use world average values~\cite{pdg98} for the branching fractions, 
$B_{\pi\pi^0} = (25.32\pm 0.15)\,\%$ and $B_e = (17.81\pm 0.07)\,\%$, 
for the $\tau$ lepton mass $M_\tau = 1777.05^{+0.29}_{-0.26}$~MeV, and 
for the CKM element $|V_{ud}| = 0.9752\pm 0.08$.  $S^e_{EW}$ represents 
the radiative corrections to leptonic $\tau$ decay, evaluated to be 
0.996~(see Ref.~\cite{MS}).  The overall radiative 
correction factor $S_{EW} = S^{\pi\pi}_{EW}/S^e_{EW}$ is estimated 
to be 1.0194~\cite{BNP}, 
based mainly on the logarithmic terms associated 
with short-distance diagrams involving loops containing bosons,  
which differ for leptonic and hadronic final states.  
Non-logarithmic terms have not yet been computed for 
the $\pi\pi$ final state, but are expected to be small.

$|F_\pi(M)|^2$ is computed from $\overline{v}(M)$ using Eq.~\ref{eq:fpi}.  
A small ($<1\%$) correction is made to represent each point as a measurement 
at the central value of its mass bin, rather than as an average over the bin.  
This is done by employing the results 
from the fits described earlier to estimate the effect of the line shape 
variation across each of the histogram bins.  Although this correction 
depends on the model used, this model-dependence is negligible relative to 
experimental errors.  
The factors appearing in Eqs.~\ref{eq:fpi} and \ref{eq:specfun} 
are also used to recast the normalization parameters determined in the fits 
described in Sec.~\ref{s-results} 
in terms of $|F_\pi(0)|^2$, as shown in Table~\ref{tab:results}.  

\begin{figure}
   \leavevmode
   \centering 
   \epsfxsize=6.5in
   \epsfbox{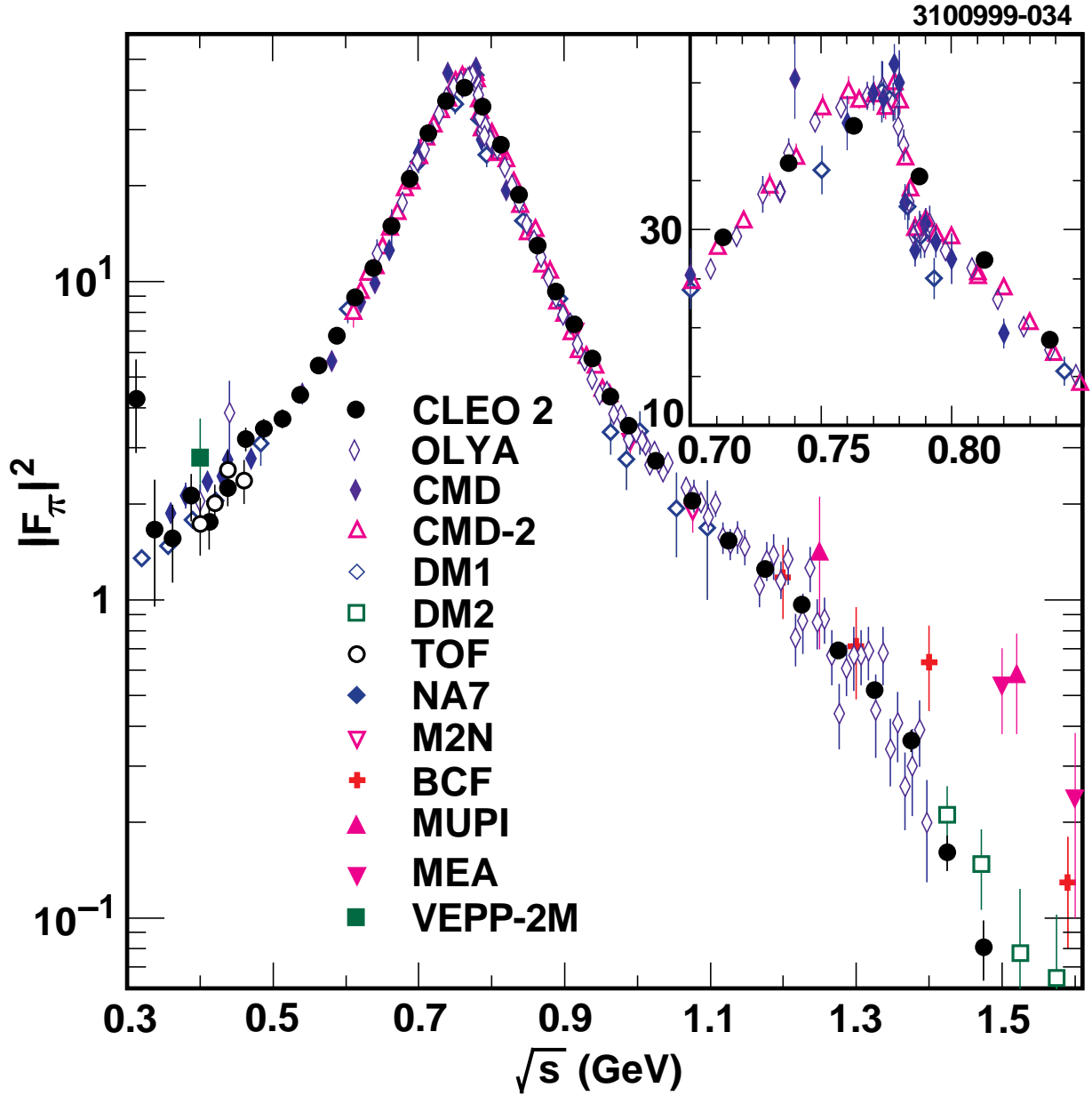}
   \caption{Comparison of $|F_\pi|^2$ as determined from CLEO-II 
            $\tau^-\to\pi^-\pi^0\nu_\tau$ data (filled circles), 
            with that obtained from $e^+e^-\to\pi^+\pi^-$ 
            cross sections (other symbols) from CMD, CMD-2, 
            DM1, DM2, OLYA, and NA7, as well as 
            Adone and other VEPP 
            experiments. 
            The inset is a blow-up of the region near the $\rho$
            peak, where $\rho$-$\omega$ interference is evident in 
            the $e^+e^-$ data.
           }
   \label{fig:cvc}
\end{figure}

\begin{figure}
   \leavevmode
   \centering 
   \epsfxsize=6.5in
   \epsfbox{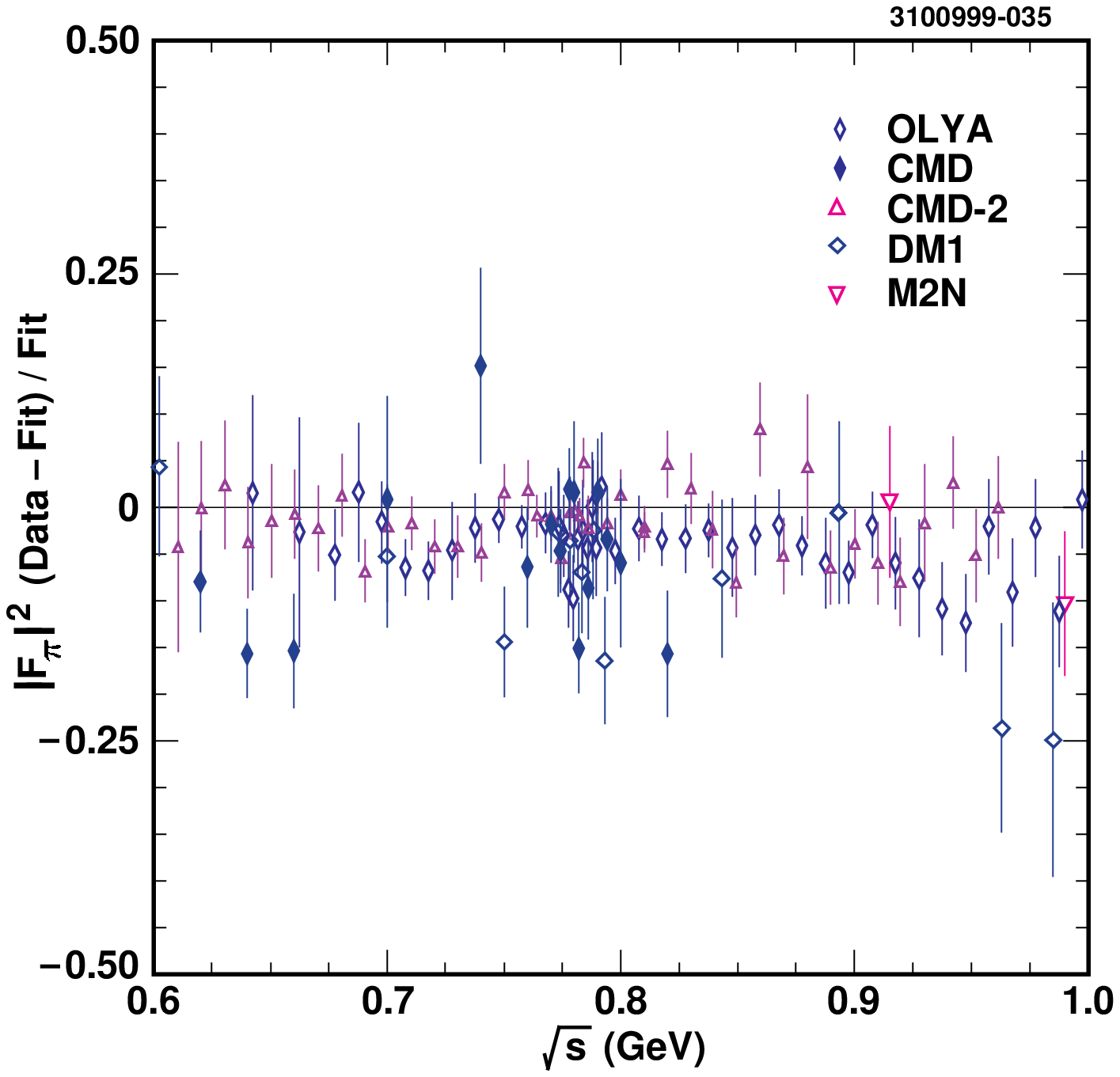}
   \caption{Difference between $|F_\pi|^2$ as determined from 
            $e^+e^-\to\pi^+\pi^-$ data and that inferred from 
            fit of the CLEO 
            $\tau^-\to\pi^-\pi^0\nu_\tau$ data to the 
            Gounaris-Sakurai model, with $\rho^{\prime\prime}$ 
            included, divided by the fit value. 
            The different symbols represent the data points 
            from the $e^+e^-$ experiments operating in the 
            $\rho$ peak region. 
           }
   \label{fig:fpidiff}
\end{figure}

In Figure~\ref{fig:cvc}, the values of $|F_\pi|^2$ derived from the 
corrected CLEO $M_{\pi^-\pi^0}$ spectrum are plotted along with the 
with $e^+e^-\to\pi^+\pi^-$ data from 
CMD~\cite{barkov}, CMD-2~\cite{cmd2}, DM1~\cite{dm1}, DM2~\cite{dm2}, 
OLYA~\cite{barkov}, NA7~\cite{na7}, as well as 
Adone~\cite{mupi,bcf,mea} experiments and other VEPP~\cite{barkov,tof}
experiments.  
The $\tau$ data follow the $e^+e^-$ data shape well, except in the
region where $\rho$-$\omega$ interference affects the $e^+e^-$ data.
However, the $\tau$ data tend to lie above the $e^+e^-$ data throughout 
most of the range in $\sqrt{s}$.  This is illustrated in 
Fig.~\ref{fig:fpidiff}, 
where we plot the fractional difference between measured 
values of $|F_\pi(s)|^2$ from the $e^+e^-$ data 
and the prediction from the G\&S fit (including $\rho^{\prime\prime}$) to 
the CLEO $\tau$ data.  To obtain 
this prediction, we modified the G\&S fit function for use with $e^+e^-$ 
data.  We allowed the normalization and a parameter associated with 
$\rho$-$\omega$ interference to float to determine the latter, and then 
readjusted the normalization to give that shown in Table~\ref{tab:results}.  

The deviation from zero in Fig.~\ref{fig:fpidiff} is consistent with the 
deviation from unity of the value of $|F_\pi(0)|^2 = 1.03$ inferred 
from the fit to the CLEO $\tau$ data.  (Fitting the $e^+e^-$ data with 
the normalization floating yields values of $|F_\pi(0)|^2 \sim 1.00$.)
It is also consistent with 
the observation made by Eidelman and Ivanchenko~\cite{eid} that the 
measured $\tau^-\to\pi^-\pi^0\nu_\tau$ branching fraction is larger 
than that expected from application of CVC to the $e^+e^-\to\pi^+\pi^-$ 
data by $3.2\pm 1.4\,\%$.  The consistency of these three indications 
of CVC violation is not accidental since they all involve (1) application  
of CVC to largely the same $e^+e^-$ data,
as well as the use of (2) 
the world average value for the $\tau$ decay 
branching fractions~\cite{pdg98} and (3) the same electroweak radiative 
correction factor~\cite{MS,BNP} to normalize the $\tau$ decay 
spectral function, as described above.  
Known sources of error in the first two of these components have been 
included in the $\pm 1.4\%$ error above.  
Recent estimates suggest that the uncertainty associated with 
additional radiative correction factors not yet computed could be as 
large as $\pm 0.4\%$ (see Refs.~\cite{adh,wmprivate}).  Deviations 
associated with isospin-violating effects are expected to be small, 
but carry an uncertainty not reflected above.

\section{IMPLICATIONS FOR THE MUON ANOMALOUS MAGNETIC MOMENT}
\label{s-amu}

\subsection{Introduction and Motivation}

The muon magnetic moment anomaly $a_\mu = (g_\mu-2)/2$ 
has an experimental value of 
$(11\,659\,230 \pm 84) \times 10^{-10}$~\cite{pdg98,bailey}.  
It receives significant 
contributions from uncalculable hadronic vacuum-polarization radiative 
corrections, estimated recently~\cite{DH98b} 
to be $a_\mu^{had} = (692.4\pm 6.2) \times 10^{-10}$.  The error on this 
contribution is the most significant source of uncertainty in the 
Standard Model prediction for $a_\mu$.  Experiment E821 at 
Brookhaven National Laboratory, currently under way, is aiming at 
a precision of $\sim 4\times 10^{-10}$ in order to be sensitive to the
weak interaction contribution at $15\times 10^{-10}$, 
as well as possible contributions from new physics.  
Improved knowledge of $a_\mu^{had}$ is important for interpreting 
experimental results on $a_\mu$.  In this section, we discuss the 
implications of our data on $\tau^-\to \pi^-\pi^0\nu_\tau$ for the 
determination of $a_\mu^{had}$, following the first such 
treatment of $\tau$ decay data, by Alemany, Davier and 
H\"ocker~\cite{adh} who used ALEPH data~\cite{aleph}.  

The value of $a_\mu^{had}$ is related to the 
$e^+e^-$ anihilation cross section to hadronic final states via 
the dispersion integral~\cite{gder}  
\begin{equation}
  a_\mu^{had} = \frac{\alpha_{em}^2(0)}{\pi}
                \int_{4M_\pi^2}^\infty{\frac{ds}{s}\,v(s)\,K(s)} ,
  \label{eq:amu}
\end{equation}
where $v(s)$ denotes the inclusive 
hadronic spectral function, and $\alpha_{em}(0)$ is the fine 
structure constant. 
$K(s)$ is the QED kernel 
\begin{equation}
  K(s) = x^2\, \left(1-\frac{x^2}{2}\right) 
       + (1+x)^2\, \left(1+\frac{1}{x^2}\right) \,
         \left( \ln{(1+x)} - x - \frac{x^2}{2} \right)
       + \left(\frac{1+x}{1-x}\right)\,x^2\,\ln{x}\, ,       
\end{equation}
where $x=(1-\beta_\mu)/(1+\beta_\mu)$, with 
$\beta_\mu = (1-4M_\mu^2/s)^{1/2}$.  
Consequently, predictions for $a_\mu^{had}$ are based primarily on 
$e^+e^-$ data.  By virtue of the form of $K(s)$ and the $1/s$ factor in 
the integrand above, measurements at low values of $\sqrt{s}$ where 
the $\pi^+\pi^-$ final state dominates are particularly significant.  

\subsection{\boldmath The Impact of $\tau$ Decay Data}

The smallness of the present error on $a_\mu^{had}$ reflects the use of 
CVC and $\tau$ decay data from ALEPH~\cite{adh} to improve the precision  
relative to that obtained based on $e^+e^-$ data alone.  Specifically, 
the authors of Ref.~\cite{adh} determine the $\pi^+\pi^-$ contribution 
over the interval $\sqrt{s} = 0.320-2.125$~GeV to be
\begin{equation}
  a_\mu^{\pi\pi}[0.320,\,2.125] = \left\{
  \begin{array}{cl}
   (495.86 \pm 12.46) \times 10^{-10} & (e^+e^- \mbox{\rm\ data\ only}) \\
   (500.81 \pm \phantom{1}6.03) \times 10^{-10}
                                      & (e^+e^- + \tau \mbox{\rm\ data})
  \end{array}\right. , 
  \label{eq:adh}
\end{equation}
where the $e^+e^-$ data did not include the CMD-2 data~\cite{cmd2} which 
were not available at that time.  The combining of $e^+e^-$ and $\tau$ data 
was carried out by averaging data points from different experiments 
prior to integrating the data.  An advantage of this procedure is that 
it properly weights the $\tau$ data more heavily in regions where it is 
more precise than the $e^+e^-$ data and vice-versa.  Accordingly, the 
improvement in precision indicated in Eq.~\ref{eq:adh} is greater than 
that obtained by determining $a_\mu^{\pi\pi}$ separately for $\tau$ 
and $e^+e^-$ data and averaging the results.   

In assessing the impact of the CLEO $\pi^-\pi^0$ spectral function on 
the value and precision of $a_\mu^{had}$, we perform the integration 
in Eq.~\ref{eq:amu} using our data alone.  Although this procedure 
does not possess the benefit described above, future changes to the central 
values or errors of external factors such as $B_{\pi\pi^0}$ or $S_{EW}$ 
can be propagated easily.  
However, we note that a careful determination of $a_\mu^{had}$ would 
combine the data from different experiments in some fashion, for example 
along the lines of the approach described in Ref.~\cite{adh}.

Details of our determination of $a_\mu^{\pi\pi}$, including the 
method and corrections applied to account for isospin violating effects 
are presented in Appendix~\ref{a-amu}.  Here, we report our results for 
the same interval in $\sqrt{s}$ as in 
Eq.~\ref{eq:adh}.  We obtain 
\begin{equation}
  a_\mu^{\pi\pi}[0.320,\, 2.125] = 
      (513.1 \pm 2.1 \pm 3.0 \pm 4.5)\times 10^{-10},  
\end{equation}
where the first error is due to statistics, the second error is 
due to experimental systematic uncertainies, and the third error 
reflects the uncertainties in externally determined quantities 
entering the calculation.  These sources of error are discussed 
in Appendix~\ref{a-amu}. 

The deviation between the CLEO result above and the result from 
Ref.~\cite{adh} given in Eq.~\ref{eq:adh} 
based on $e^+e^-\to \pi^+\pi^-$ data only is again indicative of 
disagreement in the normalizations, rather than in the shapes, of 
the $e^+e^-$ and $\tau$ spectral functions as discussed in the 
previous sections.  While the magnitude of this deviation is not 
significant on the scale of the reported errors, both it 
and the errors are greater than 
the projected precision of the BNL E821 experiment.  Continued 
efforts to precisely determine $a_\mu^{had}$ will be needed to 
help interpret the results of the Brookhaven experiment.

\section{SUMMARY}
\label{s-summary}

From a sample of approximately 87000 decays (after subtraction of 
backgrounds) 
of the type $\tau^-\to \pi^-\pi^0\nu_\tau$, we have investigated structure 
in the $\pi^-\pi^0$ invariant mass spectrum.  From fits to several models 
we have obtained parameters and relative couplings for the $\rho(770)$ and 
$\rho(1450)$ resonances.  Within the K\"uhn and Santamaria model~\cite{KS}  
with no $\rho(1700)$ contribution, the precisions on the 
$\rho(770)$ mass and width are 1.0 and 1.4 MeV, respectively.  
These precisions are comparable to those obtained by ALEPH~\cite{aleph} 
using the same $\tau$ decay mode, as well as those obtained from fits to 
low-energy $e^+e^-\to\pi^+\pi^-$ data~\cite{aleph,KS,barkov}.  

We find that a successful description of our $M_{\pi^-\pi^0}$ spectrum 
requires interfering Breit-Wigner line shapes associated with $\rho(770)$ 
and $\rho(1450)$ resonances.  Fits including additional contributions 
from the $\rho(1700)$ resonance are preferred, consistent with the 
observations of DM2~\cite{dm2} in $e^+e^-\to\pi^+\pi^-$.  
Fits of comparable quality are also obtained without a $\rho^{\prime\prime}$ 
contribution when the mass-dependence of the $\rho$ and $\rho^{\prime}$ widths 
is modified to include Blatt-Weisskopf barrier factors.  
We find that Gounaris-Sakurai~\cite{GS} type fits yield extrapolations 
to $q^2=0$ that are higher than but roughly consistent with $F_\pi(0)= 1$.   
The K\"uhn-Santamaria~\cite{KS} type fits lead to significantly higher 
values, such that imposing $F_\pi(0)=1$ can lead to biased values for 
$\rho$ and $\rho^\prime$ parameters.   

Quantitatively, the central values for the precisely determined $\rho(770)$
parameters are difficult to compare with those from fits by others.   
This is due to severe model dependence and 
the strong influence of data far from the $\rho$ peak.  Qualitatively, 
the shape of our $M_{\pi^-\pi^0}$ spectral function agrees well with 
that obtained by ALEPH~\cite{aleph}.  It also agrees well with 
the $e^+e^-\to\pi^+\pi^-$ data, supporting the applicability of CVC.  
Using our spectral function, we have employed CVC to infer the 
significant component of the hadronic contribution to the muon $g-2$ 
factor associated with the $e^+e^-\to\pi^+\pi^-$ cross section, obtaining 
$a_\mu^{\pi\pi}[0.320,\,2.125] = (513.1 \pm 5.8)\times 10^{-10}$.

However, we also 
observe indications of discrepancies between the overall normalization of 
$\tau$ and $e^+e^-$ data, as pointed out by 
Eidelman and Ivanchenko~\cite{eid}.  These appear in the model independent 
comparison of our spectrum with the $e^+e^-$ cross section measurements, 
as well as in the values for $|F_\pi(0)|^2$ inferred from fits to our 
spectrum which are larger than unity, and in our high value for 
$a_\mu^{\pi\pi}$.  Though larger than the deviations 
expected due to known sources of isospin violation, these could also arise 
from experimental errors in the $\tau$ decay branching fraction 
measurements or the normalization of the $e^+e^-\to\pi^+\pi^-$ 
cross section measurements, or from theoretical uncertainties in the 
estimates of radiative corrections, or from all of these sources.  At 
present the deviations are not significant on the scale of the reported 
errors.  We hope that new data from Novosibirsk, BEPC, 
and the B-factory (PEP-II, KEK-B and CESR-III) storage rings will shed 
light on this issue in the near future.  


\section*{ACKNOWLEDGEMENTS}
We acknowledge useful conversations with 
M.~Davier, S.~Eidelman, A.~H\"ocker, W.~Marciano and A.~Vainshtein.
We gratefully acknowledge the effort of the CESR staff in providing us with
excellent luminosity and running conditions.
J.R. Patterson and I.P.J. Shipsey thank the NYI program of the NSF, 
M. Selen thanks the PFF program of the NSF, 
M. Selen and H. Yamamoto thank the OJI program of DOE, 
J.R. Patterson, K. Honscheid, M. Selen and V. Sharma 
thank the A.P. Sloan Foundation, 
M. Selen and V. Sharma thank the Research Corporation, 
F. Blanc thanks the Swiss National Science Foundation, 
and H. Schwarthoff and E. von Toerne thank 
the Alexander von Humboldt Stiftung for support.  
This work was supported by the National Science Foundation, the
U.S. Department of Energy, and the Natural Sciences and Engineering Research 
Council of Canada.

\appendix

\section{UNFOLDING OF RESOLUTION AND RADIATIVE EFFECTS}
\label{a-unfold}

The unfolding procedure is used to correct for bin migration 
in the observed $M_{\pi\pi^0}$ spectrum due to resolution smearing 
and radiative effects.  In this Appendix, we describe the 
procedure used in this analysis.  
In the discussion that follows, matrices are denoted in boldface, while 
non-bold symbols represent vectors with the contents of binned 
histograms.

The unfolding procedure makes use of the Monte Carlo to characterize the 
bin migration.  One may construct a migration matrix {\boldmath $P$} 
which gives the probability that an event generated with a given 
$\pi^-\pi^0(\gamma)$ mass is reconstructed with a given $\pi^-\pi^0$ mass:
\begin{equation}
  R^{\rm MC} = \mbox{\boldmath $P$}\, G^{\rm MC}, 
\end{equation}
where $G^{\rm MC}$ represents the generated $\pi^-\pi^0(\gamma)$ mass spectrum 
and $R^{\rm MC}$ represents the reconstructed one.  It is possible to apply 
the inverse of {\boldmath $P$} to the spectrum observed in the data 
$R^{\rm data}$ to obtain an 
unfolded spectrum $U^{\rm data}$ that provides an estimate for the parent 
spectrum.  However, such a procedure is not robust with respect to statistical 
fluctuations entering the determination of {\boldmath $P$}, and can yield 
spectra with unphysically large point-by-point fluctuations.  

The corrected ALEPH 
spectrum~\cite{aleph} was derived using a method based on singular value 
decomposition of the migration matrix~\cite{hocker} to mitigate this 
effect.\footnote{For fits to models, ALEPH performed 
fits of their uncorrected spectrum to convolutions of the experimental 
effects with the functions describing the models, avoiding this 
problem altogether.}
Here, we use an iterative method that relies on the smallness 
of the bin migrations and the approximate similarity between the 
reconstructed spectra from the data ($R^{\rm data}$) and 
Monte Carlo ($R^{\rm MC}$) samples.  
We construct the matrix {\boldmath $P^\prime$}, which gives the fraction 
of MC events with a given reconstructed $\pi^-\pi^0$ mass that were generated 
with a given $\pi^-\pi^0(\gamma)$ mass.  With this matrix, 
$G^{\rm MC} = \mbox{\boldmath $P^\prime$}\, R^{\rm MC}$ is satisfied, but 
{\boldmath $P^\prime$} is not equal to {\boldmath $P^{-1}$}.  Successive 
application of {\boldmath $P^{\prime}$} to the observed data spectrum 
gives an estimate for the parent distribution, according to:
\begin{equation}
        U^{\rm data}  =  \mbox{\boldmath $P^{\prime}$}\, R^{\rm data}
        + \sum_{k=1}^\infty{({\bf 1} - \mbox{\boldmath $P^{\prime} P$})^k\, 
                                       \mbox{\boldmath $P^\prime$}\, 
                                       (R^{\rm data} - R^{\rm MC}) }.
\label{eq:unfold}
\end{equation}
For small bin migration probabilities, all elements of the matrix 
$({\bf 1} - \mbox{\boldmath $P^\prime P$})$ are small: these elements are 
the `expansion parameters' in the series above.  A simpler form for 
Eq.~\ref{eq:unfold} is obtained by recognizing that the quantity 
$({\bf 1} - \mbox{\boldmath $P^\prime P$})\mbox{\boldmath $P^\prime$}\, 
R^{\rm MC}$ 
is identically zero.  As written, however, 
Eq.~\ref{eq:unfold} illustrates that 
when the series is truncated, deviation from the results of the full 
expansion vanishes as $(R^{\rm data} - R^{\rm MC}) \rightarrow 0$.  

With the similarity between the observed data and MC spectra plotted 
in Fig.~\ref{fig:mpipi}, we find it sufficient to ignore terms with $k>1$ 
in Eq.~\ref{eq:unfold}.  The $k=1$ term has a noticeable effect on 
several of the fit parameters reported in Section~\ref{ss-ksfit}.  In 
particular, ignoring this term leads to a value for $\Gamma_\rho$ that
is 0.5 MeV smaller than that from our nominal fit.  Higher order terms 
have no significant impact on any of the parameters.

\section{\boldmath THE CORRECTED $\pi^-\pi^0$ MASS SPECTRUM}
\label{a-massspectrum}

In this appendix we present the fully corrected CLEO $M_{\pi\pi^0}$ 
spectrum in tabular form.  The spectrum is given as a compilation 
of event yields for each mass bin in Table~\ref{tab:cleocorr}, 
normalized so that the sum of entries over all bins is unity.  Also 
given are the square roots of the diagonal elements of the covariance 
matrix (statistical errors only).  
\begin{table}
\caption[]{The corrected CLEO $M_{\pi\pi^0}$ spectrum given in terms 
           of event yield as a function of mass bin.  The spectrum 
           is normalized so that the number of entries sums to unity.
           The numbers in parentheses denote the square roots of the 
           diagonal elements of the covariance matrix (statistical 
           errors only).  Note that the number of entries jumps at 
           the 1.00--1.05~GeV bin, where the bin size is increased 
           from 0.025 to 0.050~GeV.
          }
\label{tab:cleocorr}
\begin{tabular}{rcc c rcc}
  Bin & Mass Range &  Entries &  \phantom{0} &
  Bin & Mass Range &  Entries \\ 
  No. & (GeV)      &  $(10^{-4})$  &   &
  No. & (GeV)      &  $(10^{-4})$  \\ 
\hline
  1 & 0.275--0.300 & $   1.5\;  (1.4)$ &   & & & \\
  2 & 0.300--0.325 & $   8.0\;  (2.5)$ &   & & & \\
  3 & 0.325--0.350 & $   6.0\;  (2.6)$ &   & & & \\
  4 & 0.350--0.375 & $   8.5\;  (2.3)$ &   & 24 & 0.850--0.875 & $ 446.7\;  (6.6)$ \\
  5 & 0.375--0.400 & $  15.6\;  (2.6)$ &   & 25 & 0.875--0.900 & $ 326.2\;  (5.5)$ \\
  6 & 0.400--0.425 & $  16.2\;  (2.9)$ &   & 26 & 0.900--0.925 & $ 262.1\;  (4.9)$ \\
  7 & 0.425--0.450 & $  24.9\;  (3.1)$ &   & 27 & 0.925--0.950 & $ 207.3\;  (4.3)$ \\
  8 & 0.450--0.475 & $  41.4\;  (3.4)$ &   & 28 & 0.950--0.975 & $ 158.8\;  (3.7)$ \\
  9 & 0.475--0.500 & $  50.6\;  (3.7)$ &   & 29 & 0.975--1.000 & $ 129.6\;  (3.4)$ \\
 10 & 0.500--0.525 & $  60.9\;  (4.0)$ &   & 30 & 1.000--1.050 & $ 202.8\;  (4.8)$ \\
 11 & 0.525--0.550 & $  79.8\;  (4.4)$ &   & 31 & 1.050--1.100 & $ 151.0\;  (4.1)$ \\
 12 & 0.550--0.575 & $ 107.4\;  (4.7)$ &   & 32 & 1.100--1.150 & $ 111.0\;  (3.4)$ \\
 13 & 0.575--0.600 & $ 144.3\;  (5.2)$ &   & 33 & 1.150--1.200 & $  87.0\;  (3.0)$ \\
 14 & 0.600--0.625 & $ 204.5\;  (5.9)$ &   & 34 & 1.200--1.250 & $  63.9\;  (2.5)$ \\
 15 & 0.625--0.650 & $ 269.1\;  (6.5)$ &   & 35 & 1.250--1.300 & $  42.7\;  (2.0)$ \\
 16 & 0.650--0.675 & $ 385.8\;  (7.5)$ &   & 36 & 1.300--1.350 & $  29.2\;  (1.7)$ \\
 17 & 0.675--0.700 & $ 571.5\;  (8.7)$ &   & 37 & 1.350--1.400 & $  18.1\;  (1.3)$ \\
 18 & 0.700--0.725 & $ 826.8\; (10.1)$ &   & 38 & 1.400--1.450 & $  6.98\; (0.84)$ \\
 19 & 0.725--0.750 & $1078.4\; (11.3)$ &   & 39 & 1.450--1.500 & $  2.91\; (0.59)$ \\
 20 & 0.750--0.775 & $1228.1\; (11.8)$ &   & 40 & 1.500--1.550 & $  0.71\; (0.32)$ \\
 21 & 0.775--0.800 & $1114.7\; (11.0)$ &   & 41 & 1.550--1.600 & $  0.59\; (0.25)$ \\
 22 & 0.800--0.825 & $ 878.1\;  (9.6)$ &   & 42 & 1.600--1.650 & $  0.68\; (0.26)$ \\
 23 & 0.825--0.850 & $ 629.3\;  (7.9)$ &   & 43 & 1.650--1.700 & $  0.28\; (0.21)$ \\
\end{tabular}
\end{table}

The statistical errors given in Table~\ref{tab:cleocorr} do not reflect 
the correlations between data points that were introduced by the bin 
migration correction.  Although the correlations are not 
large, proper treatment of these data necessitates use of the 
covariance matrix.  The full covariance matrix $V$ is a $43\times 43$ 
symmetric matrix.  In Table~\ref{tab:covar}, we present the correlation 
coefficients $\rho_{ij} = V_{ij}/(V_{ii}V_{jj})^{1/2}$ for bins $i$ and 
$j$, $i>j$, for $i,\,j$ where $\rho_{ij}>0.0015$.  The coefficients shown 
are the statistical correlations only.  Both the corrected spectrum 
as given in Table~\ref{tab:cleocorr} and the full covariance matrix are 
available electronically~\cite{wwwrho}.  
\begin{table}
\caption[]{Correlation coefficients $\rho_{ij}$ for contents of 
           bins $i$ and $j$ (see Table~\ref{tab:cleocorr}) 
           of the fully corrected CLEO $M_{\pi\pi^0}$
           spectrum, for $i>j$.  Coefficients with values less than 0.0015 
           are not shown.  These coefficients reflect statistical 
           correlations only.  
          }
\label{tab:covar}
\begin{tabular}{rrr c rrr c rrr c rrr c rrr c rrr}
 $i$ & $j$ & $\rho_{ij}$ & \phantom{.} &
 $i$ & $j$ & $\rho_{ij}$ & \phantom{.} &
 $i$ & $j$ & $\rho_{ij}$ & \phantom{.} &
 $i$ & $j$ & $\rho_{ij}$ & \phantom{.} &
 $i$ & $j$ & $\rho_{ij}$ & \phantom{.} &
 $i$ & $j$ & $\rho_{ij}$ \\
\hline 
  2 &  1 & $-.002$ & & 10 &  8 & $-.057$ & & 16 & 14 & $-.084$ & &
 21 & 19 & $-.116$ & & 28 & 27 & $ .299$ & & 38 & 35 & $-.011$ \\
  3 &  1 & $-.031$ & & 10 &  9 & $ .085$ & & 16 & 15 & $ .183$ & &
 21 & 20 & $ .205$ & & 29 & 25 & $-.006$ & & 38 & 36 & $-.074$ \\
  3 &  2 & $ .042$ & & 11 &  6 & $-.002$ & & 17 &  3 & $ .002$ & &
 22 & 18 & $-.003$ & & 29 & 26 & $-.046$ & & 38 & 37 & $ .114$ \\
  4 &  1 & $ .003$ & & 11 &  8 & $-.013$ & & 17 & 11 & $-.002$ & &
 22 & 19 & $-.033$ & & 29 & 27 & $-.102$ & & 39 & 35 & $ .002$ \\
  4 &  2 & $-.025$ & & 11 &  9 & $-.064$ & & 17 & 12 & $-.003$ & &
 22 & 20 & $-.113$ & & 29 & 28 & $ .317$ & & 39 & 36 & $-.009$ \\
  4 &  3 & $ .053$ & & 11 & 10 & $ .120$ & & 17 & 13 & $-.003$ & &
 22 & 21 & $ .219$ & & 30 & 26 & $-.004$ & & 39 & 37 & $-.082$ \\
  5 &  1 & $ .007$ & & 12 &  2 & $ .002$ & & 17 & 14 & $-.023$ & &
 23 &  1 & $ .003$ & & 30 & 27 & $-.034$ & & 39 & 38 & $ .103$ \\
  5 &  2 & $-.008$ & & 12 &  7 & $-.003$ & & 17 & 15 & $-.090$ & &
 23 & 19 & $-.003$ & & 30 & 28 & $-.094$ & & 40 & 37 & $-.015$ \\
  5 &  3 & $-.034$ & & 12 &  9 & $-.014$ & & 17 & 16 & $ .201$ & &
 23 & 20 & $-.036$ & & 30 & 29 & $ .123$ & & 40 & 38 & $-.079$ \\
  5 &  4 & $ .012$ & & 12 & 10 & $-.076$ & & 18 & 13 & $-.002$ & &
 23 & 21 & $-.107$ & & 31 & 28 & $-.019$ & & 40 & 39 & $ .166$ \\
  6 &  1 & $-.004$ & & 12 & 11 & $ .145$ & & 18 & 14 & $-.005$ & &
 23 & 22 & $ .241$ & & 31 & 29 & $-.071$ & & 41 & 38 & $-.011$ \\
  6 &  2 & $ .002$ & & 13 &  9 & $-.002$ & & 18 & 15 & $-.025$ & &
 24 & 20 & $-.003$ & & 31 & 30 & $ .029$ & & 41 & 39 & $-.090$ \\
  6 &  3 & $-.005$ & & 13 & 10 & $-.017$ & & 18 & 16 & $-.096$ & &
 24 & 21 & $-.037$ & & 32 & 29 & $-.007$ & & 41 & 40 & $ .152$ \\
  6 &  4 & $-.031$ & & 13 & 11 & $-.075$ & & 18 & 17 & $ .206$ & &
 24 & 22 & $-.104$ & & 32 & 30 & $-.061$ & & 42 & 36 & $-.002$ \\
  6 &  5 & $ .027$ & & 13 & 12 & $ .150$ & & 19 & 12 & $-.002$ & &
 24 & 23 & $ .255$ & & 32 & 31 & $ .046$ & & 42 & 39 & $-.013$ \\
  7 &  4 & $-.006$ & & 14 &  3 & $ .002$ & & 19 & 14 & $-.002$ & &
 25 & 21 & $-.003$ & & 33 & 30 & $-.005$ & & 42 & 40 & $-.094$ \\
  7 &  5 & $-.041$ & & 14 &  9 & $-.002$ & & 19 & 15 & $-.004$ & &
 25 & 22 & $-.039$ & & 33 & 31 & $-.066$ & & 42 & 41 & $ .226$ \\
  7 &  6 & $ .059$ & & 14 & 10 & $-.002$ & & 19 & 16 & $-.028$ & &
 25 & 23 & $-.102$ & & 33 & 32 & $ .056$ & & 43 & 18 & $ .002$ \\
  8 &  2 & $-.002$ & & 14 & 11 & $-.019$ & & 19 & 17 & $-.103$ & &
 25 & 24 & $ .261$ & & 34 & 31 & $-.006$ & & 43 & 30 & $-.002$ \\
  8 &  4 & $-.002$ & & 14 & 12 & $-.078$ & & 19 & 18 & $ .207$ & &
 26 & 22 & $-.004$ & & 34 & 32 & $-.071$ & & 43 & 31 & $ .006$ \\
  8 &  5 & $-.009$ & & 14 & 13 & $ .155$ & & 20 & 13 & $-.002$ & &
 26 & 23 & $-.041$ & & 34 & 33 & $ .068$ & & 43 & 35 & $-.003$ \\
  8 &  6 & $-.053$ & & 15 & 11 & $-.003$ & & 20 & 14 & $-.002$ & &
 26 & 24 & $-.104$ & & 35 & 32 & $-.007$ & & 43 & 36 & $ .007$ \\
  8 &  7 & $ .067$ & & 15 & 12 & $-.019$ & & 20 & 15 & $-.002$ & &
 26 & 25 & $ .270$ & & 35 & 33 & $-.073$ & & 43 & 37 & $-.002$ \\
  9 &  3 & $-.003$ & & 15 & 13 & $-.082$ & & 20 & 16 & $-.003$ & &
 27 & 23 & $-.004$ & & 35 & 34 & $ .074$ & & 43 & 40 & $-.023$ \\
  9 &  4 & $-.003$ & & 15 & 14 & $ .178$ & & 20 & 17 & $-.030$ & &
 27 & 24 & $-.042$ & & 36 & 33 & $-.009$ & & 43 & 41 & $-.111$ \\
  9 &  6 & $-.010$ & & 16 &  2 & $ .003$ & & 20 & 18 & $-.111$ & &
 27 & 25 & $-.106$ & & 36 & 34 & $-.076$ & & 43 & 42 & $ .246$ \\
  9 &  7 & $-.056$ & & 16 &  3 & $-.002$ & & 20 & 19 & $ .200$ & &
 27 & 26 & $ .276$ & & 36 & 35 & $ .076$ & &  & & \\
  9 &  8 & $ .075$ & & 16 & 10 & $-.002$ & & 21 & 16 & $-.002$ & &
 28 & 24 & $-.004$ & & 37 & 34 & $-.008$ & &  & & \\
 10 &  5 & $-.002$ & & 16 & 12 & $-.003$ & & 21 & 17 & $-.003$ & &
 28 & 25 & $-.044$ & & 37 & 35 & $-.073$ & &  & & \\
 10 &  7 & $-.011$ & & 16 & 13 & $-.022$ & & 21 & 18 & $-.032$ & &
 28 & 26 & $-.103$ & & 37 & 36 & $ .081$ & &  & & \\
\end{tabular}
\end{table}

\section{DETERMINATION OF $\mbox{\boldmath $a_\mu^{\pi\pi}$}$ 
         FROM CLEO DATA}
\label{a-amu}

In this appendix, we present the details of our determination of 
$a_\mu^{\pi\pi}$ over the range $\sqrt{s}= 0.320$ to 2.125 GeV.
Quantities entering this determination are summarized in 
Table~\ref{tab:amu}.  We describe the integration procedure, 
corrections for isospin-violating effects, and the evaluation of 
errors below.  Finally we discuss the implications of our results.
\begin{table}[hbt]
\caption{Components of the derivation of $a_\mu^{\pi\pi}[0.320,\, 2.125]$ 
         from the CLEO $\tau^-\to\pi^-\pi^0\nu_\tau$ hadronic mass 
         spectrum.  
        }
\label{tab:amu}
  \begin{tabular}{lccc}
  Quantity  & Value & Correction to                 & Uncertainty on \\
            &       & $a_\mu^{\pi\pi}$ ($10^{-10}$) & 
                      $a_\mu^{\pi\pi}$ ($10^{-10}$) \\ 
\hline
  \multicolumn{4}{l}{Integration of $v(s)$ (Eq.~\ref{eq:amu}) }\\
   $a_\mu^{\pi\pi}[0.320,\, 2.125]$ (raw)  
           & $(514.8\pm 2.1)\times 10^{-10}$ & --- & $\pm 2.1$ \\
\hline
  \multicolumn{4}{l}{Normalization Factors} \\
  $B(\tau^-\to\pi^-\pi^0\nu_\tau)$~\cite{pdg98} 
           & $(25.32\pm 0.15)\,\%$ & --- & $\pm 3.0$ \\
  $B(\tau^-\to e^-\overline{\nu}\,\!_e\nu_\tau)$~\cite{pdg98}
           & $(17.81\pm 0.07)\,\%$ & --- & $\pm 2.0$ \\
  $S_{EW}$~\cite{BNP}   & $1.0194\pm 0.0040$  & --- & $\pm 2.0$ \\
  $V_{ud}$~\cite{pdg98} & $0.9752\pm 0.0008$  & --- & $\pm 0.8$ \\
\hline
  \multicolumn{4}{l}{Correction Factors} \\
  $\alpha$ ($\rho$-$\omega$ interference) 
           & $(1.71\pm 0.06\pm 0.20)\times 10^{-3}$  
           & $+3.6$ & $\pm 0.4$ \\
  $\Delta\Gamma_\rho$~\cite{adh} 
           & $(+0.28\pm 0.39)\,\%$ & $+1.1$ & $\pm 1.6$ \\
  $M_{\pi^-}-M_{\pi^0}$ 
           & 4.6 MeV & $-6.5$ & --- \\
\hline
  \multicolumn{4}{l}{Other Sources of Systematic Error} \\
  Backgrounds              & ---         & --- & $\pm 2.4$ \\
  Bin Migration            & ---         & --- & $\pm 1.4$ \\
  Energy Scale             & $\pm 0.3\%$ & --- & $\pm 1.0$ \\
  Acceptance               & ---         & --- & $\pm 0.5$ \\
  Integration Procedure    & ---         & --- & $\pm 1.0$ \\
  \end{tabular}
\end{table}

\subsection{Integration Procedure}

To evaluate $a_\mu^{\pi\pi}$ as given by Eq.~\ref{eq:amu}, we perform 
a numerical integration employing the Gounaris-Sakurai model, 
with $\rho^{\prime\prime}$ included, using the 
best fit parameters given in Table~\ref{tab:results}.  The externally 
measured quantities used to infer $v^{\pi\pi^0}$ from our corrected 
$\pi^-\pi^0$ mass spectrum are listed in Table~\ref{tab:amu}. 
From this procedure we obtain, 
prior to application of the corrections described below, 
\begin{equation}
  a_\mu^{\pi\pi}[0.320,\, 2.125] = (514.8 \pm 2.1) \times 10^{-10}, 
\end{equation}
where the error is statistical only.

The advantages of this procedure 
relative to direct integration of the data points (summing the histogram 
entries) are (1) mitigation of the effects of statistical fluctuations 
particularly in the low-mass bins and (2) operational simplicity.  
The disadvantages include possible biases associated with choice of 
model.  We have checked this method of integrating the functional 
form of $v^{\pi\pi}(s)$ by reproducing the CMD-2 evaluation~\cite{cmd2} 
of $a_\mu^{\pi\pi}[0.61,\,0.96]$ to within $0.6\times 10^{-10}$ of 
the value they obtained via direct integration of their data.  We 
have also verified with our data that direct integration gives comparable 
results to those obtained by integrating the fit function.  

\subsection{Corrections}

Several corrections are needed to account for sources of isospin 
violation that bias the naive application of CVC.  The corrections 
can be classified according to three 
quantities: (1) the magnitude of the $\rho$-$\omega$ interference 
arising from the isospin-violating electromagnetic decay 
$\omega \to\pi^+\pi^-$; (2) possible isospin splittings between 
charged and neutral $\rho$ meson masses and widths; and (3) 
kinematic effects associated with the $\pi^-$/$\pi^0$ mass 
difference.  These corrections are listed in Table~\ref{tab:amu}, 
and are described below.

\subsubsection{Contributions from $\omega\to\pi^+\pi^-$}

To account for the absence of $\rho$-$\omega$ interference in 
$\tau$ data, we modify the G\&S function to include it, introducing 
the parameter $\alpha$ (following the notation of Ref.~\cite{KS}) to 
quantify the $\omega$ admixture, analogous to the parameters 
$\beta$ and $\gamma$ which quantify the $\rho^{\prime}$ and 
$\rho^{\prime\prime}$ amplitudes relative to that of the $\rho$ meson. 
From fits to the $e^+e^-$ data, we find 
$\alpha = (1.71\pm 0.06\pm 0.20)\times 10^{-3}$.  Modifying our 
fit function in this way leads to an increase in 
$a_\mu^{\pi\pi}$ by $3.6\times 10^{-10}$ units.

\subsubsection{$\rho$ meson isospin splittings}

Following the authors of Ref.~\cite{adh}, the charged-neutral $\rho$ 
mass splitting, expected to be small, is taken to be zero. 
We use their evaluation~\cite{adh} of the pole width splitting
$\Delta\Gamma_\rho = (\Gamma_{\rho^-} - \Gamma_{\rho^0})/\Gamma_\rho 
= (2.8\pm 3.9)\times 10^{-3}$.  The dominant source of 
this splitting is the $\pi^-/\pi^0$ mass difference, which gives rise 
to different kinematic factors for charged and neutral $\rho$ decay 
(see Eq.~\ref{eq:gamma}).  Additional small differences 
between charged and neutral $\rho$ meson decay also affect the widths, 
some of which are included in the above estimate.  One possibly 
significant difference, not accounted for here, is the effect of 
final state Coulomb interactions in $\rho^0\to \pi^+\pi^-$ decay 
which would tend to enhance the $\rho^0$ width~\cite{wmprivate}.  
Finally, as in Ref.~\cite{adh}, 
we assume that the charged and neutral $\rho^\prime$ and 
$\rho^{\prime\prime}$ parameters are the same.  

In determining the effect of the estimated $\Delta \Gamma_\rho$ on 
$a_\mu^{\pi\pi}$, we modify $\Gamma_\rho$ as it appears in the 
denominator of the expression for $v^{\pi\pi}(q^2)$ (see, for example, 
Eq.~\ref{eq:bw}).  In the context of the K\&S and G\&S models, where 
the $F_\pi(0)=1$ constraint is enforced, 
the $\Gamma_\rho$ factor does not appear explicitly 
in the numerator of the squared Breit-Wigner formula, 
unlike the general form for $v^{\pi\pi}(q^2)$ given by Eq.~\ref{eq:bw}.
Modifying $\Gamma_\rho$ in the denominator only 
leads to an additive correction to 
$a_\mu^{\pi\pi}$ of $(+1.1\pm 1.6)\times 10^{-10}$.  This is contrary 
to the result of Ref.~\cite{adh}, in which the correction is given as 
$-1.4 \times 10^{-10}$.  

\subsubsection{Additional kinematic factors}

As mentioned above, the $\pi^-/\pi^0$ mass difference contributes 
to $\Delta \Gamma_\rho$ through the $P$-wave phase space factor 
$(2p_\pi/\sqrt{q^2})^3$ appearing in Eq.~\ref{eq:gamma}.  This 
factor also characterizes the $q^2$ dependence of the $\rho$ width 
and affects the numerator as well as the denominator of 
$v^{\pi\pi}(q^2)$ in the models considered.  This effect 
influences the spectral function strongly at low values of $q^2$, 
since that is where the values of $p_\pi$ for charged and neutral 
$\rho$ meson decay differ most significantly.  
Accounting for this difference leads to a decrease in $a_\mu^{\pi\pi}$ 
by $6.5\times 10^{-10}$.  

\subsection{Errors}

In this section, we discuss the sources of error 
indicated in Table~\ref{tab:amu}. 

\subsubsection{Statistical Errors} 

Since we use external measurements to normalize our spectral function, 
the statistical errors considered here are those associated with the 
bin-by-bin fluctuations in our mass spectrum.  Statistical errors associated 
with the Monte Carlo based corrections for backgrounds, bin migration and 
acceptance also enter.  We assess the overall statistical error by generating 
a large number of G\&S parameter sets, with the parameters determined 
randomly about the central values returned by our nominal fit, weighted 
according to the covariance matrix returned by the fit, assuming Gaussian 
errors.  We determine $a_\mu^{\pi\pi}$ separately for each parameter set.  
The r.m.s. of the distribution of values was found to be $2.1\times 10^{-10}$. 

\subsubsection{Internal Systematic Errors}

Internal systematic errors are those associated with our analysis of 
$\tau^-\to\pi^-\pi^0\nu_\tau$ decays.  They originate from the sources 
indicated in Sec.~\ref{s-syserr} in the context of our fits to models 
of the $\pi^-\pi^0$ mass spectrum.  As with the statistical error, these 
errors pertain to the shape, rather than the normalization, of the 
spectral function.  

As expected from Table~\ref{tab:syserr}, 
the dominant sources are uncertainties associated 
with the background subtraction and bin migration corrections.  Possible 
biases in these corrections would tend to affect the low end and $\rho$ 
peak regions of the mass spectrum, on which $a_\mu^{\pi\pi}$ 
depends most sensitively.  We have also considered 
energy scale and acceptance uncertainties.  
We have estimated the uncertainties associated with these sources, 
shown in Table~\ref{tab:amu}, in the same ways as described 
in Sec.~\ref{s-syserr}.  

We have also estimated the bias associated with the model dependence 
of the approach used to compute $a_\mu^{\pi\pi}$.  This has been done 
by comparing values of $a_\mu^{\pi\pi}$ obtained with different models, 
as well as by directly integrating the data points.  We estimate an 
uncertainty of $\pm 1.0\times 10^{-10}$ from this source.  Adding this 
in quadrature with the errors described above yields an 
overall internal systematic error of $\pm 3.0 \times 10^{-10}$.  

\subsubsection{External Systematic Errors} 

External systematic errors are those associated with the parameters used 
to infer $a_\mu^{\pi\pi}$ from our corrected $\pi\pi^0$ mass spectrum.  
They include uncertainties associated with normalization factors, of which 
$B_{\pi\pi^0}$, $B_e$, $S_{EW}$, and $V_{ud}$ contribute the dominant 
errors.  They also include the uncertainties associated with the 
corrections for isospin-violating effects described in the previous 
section.  Adding the errors listed in Table~\ref{tab:amu} for these 
sources in quadrature gives an overall external systematic error of 
$\pm 4.5 \times 10^{-10}$.

\subsection{Results and Discussion}

With the normalization and correction factors listed in Table~\ref{tab:amu}, 
we obtain
\begin{eqnarray}
  a_\mu^{\pi\pi}[0.320,\, 2.125] & = &
      (513.1 \pm 2.1 \pm 3.0 \pm 4.5)\times 10^{-10}  \nonumber \\
     & &  \times \left( \frac{B_{\pi\pi^0}}{0.2532} \right)
          \times \left( \frac{0.1781}{B_e} \right)
          \times \left( \frac{1.0194}{S_{EW}} \right)
          \times \left( \frac{0.9752}{|V_{ud}|} \right) ^2 \, , 
\end{eqnarray}
where the first error is the statistical error, the second is the 
internal systematic error, and the third is the external systematic error.  
In the above expression, we have made explicit the dependence on the 
external normalization factors so as to facilitate incorporation of 
future measurements of these quantities.  

Since this evaluation of $a_\mu^{\pi\pi}$ is independent of the $e^+e^-$ 
only estimate from Ref.~\cite{adh}, we can perform a weighted average 
of the two results.  From this, we obtain
\begin{equation}
  a_\mu^{\pi\pi}[0.320,\, 2.125] = (510.0 \pm 5.3) \times 10^{-10}.  
\end{equation}
This determination does not include the ALEPH data~\cite{aleph}, nor 
does it include the CMD-2 data~\cite{cmd2}.  This is larger than the 
value of $(500.81\pm 6.03)\times 10^{-10}$ obtained by the authors of 
Ref.~\cite{adh}, despite the apparent agreement of the CLEO and ALEPH 
$\pi^-\pi^0$ mass spectra.  This reflects in part the difference in the 
procedures of combining the data, as described above.  

%
%
%

%
%

\begin{references}
\bibitem{chrg}       Generalization to charge conjugate reactions 
                       and states is implied throughout, except as noted.
\bibitem{dsinv}      The inverse of the covariance matrix is computed 
                       using the {\tt DSINV} symmetric real matrix inversion 
                       routine from the CERN program library.
\end{references}
\end{document}